\documentclass[graybox, envcountchap]{svmult}

\usepackage{mathptmx}        % selects Times Roman as basic font
\usepackage{amsmath}
\usepackage{amssymb,comment}
\usepackage{color}
\usepackage{helvet}          % selects Helvetica as sans-serif font
\usepackage{courier}         % selects Courier as typewriter font
\usepackage{dirtree}
\usepackage{makeidx}        % allows index generation
\usepackage{graphicx}        % standard LaTeX graphics tool
\usepackage{subfig}

\usepackage{multicol}        % used for the two-column index
\usepackage[bottom]{footmisc}% places footnotes at page bottom

\usepackage{hyperref}        %for hyperlinks
\hypersetup{colorlinks=true,urlcolor=blue}

\usepackage[misc]{ifsym}

\usepackage[numbers]{natbib}
\makeindex             % used for the subject index
\usepackage{xspace}
\newcommand{\nustar}{\textit{NuSTAR}\xspace}
\newcommand{\ixpe}{\textit{IXPE}\xspace}
\newcommand{\rxte}{\textit{RXTE}\xspace}

\newcommand{\integral}{\textit{INTEGRAL}\xspace}

\newcommand{\mdotw}{$\dot{M}_w$\xspace}
\newcommand{\xrism}{\textit{XRISM}\xspace}
\newcommand{\athena}{\textit{Athena}\xspace}
\newcommand{\chandra}{\textit{Chandra}\xspace}
\newcommand{\xmm}{\textit{XMM-Newton}\xspace}

\begin{document}

%%%%%%%%%%%%%%%%%%%%%%%%%%%%%%%%%%%%%%%%%%%%%%%%%%%%%%%%%%%%%%%%%

\setcounter{chapter}{12}
\title{High-Resolution Spectroscopy of X-ray Binaries}

\author{Joey Neilsen and Nathalie Degenaar}

\institute{Villanova University, Mendel Science Center 263B, 800 E. Lancaster Ave., Villanova PA 19085, USA, \email{jneilsen@villanova.edu} (corresponding author)\\\\
Anton Pannekoek Institute, University of Amsterdam, Postbus 94249, 1090 GE Amsterdam, NL; \email{N.D.Degenaar@uva.nl} (corresponding author)}

\maketitle

\abstract{X-ray binaries, as bright local sources with short variability timescales for a wide range of accretion processes, represent ideal targets for high-resolution X-ray spectroscopy. In this chapter, we present a high-resolution X-ray spectral perspective on X-ray binaries, focusing on black holes and neutron stars. The majority of the chapter is devoted to observational and theoretical signatures of mass ejection via accretion disk winds: we discuss their appearance (including an overview of photoionization and thermodynamic processes that determine their visibility in X-ray spectra) and their life cycles (including efforts to constrain their time-dependent mass loss rates), and we provide a broad overview of the primary accretion disk wind driving mechanisms that have been considered in the literature: (1) radiation pressure, where radiation accelerates a wind by scattering off electrons or atoms in the disk or its atmosphere; (2) thermal driving, where Compton heating of the outer accretion disk causes gas thermal velocities to exceed the local escape speed; and (3) magnetohydrodynamic processes, where gas may be ejected from the disk via magnetic pressure gradients or magnetocentrifugal effects. We then turn to spectroscopic constraints on the geometry of accreting systems, from relativistically blurred emission lines to dipping sources, clumpy, structured stellar winds, and baryonic jets. We conclude with discussions of measurements of the interstellar medium and the potential of next-generation high-resolution X-ray spectroscopy for X-ray binaries.\vspace{60pt}}

\section{Introduction}
X-ray binaries (XRBs) consist of a compact object (white dwarf, neutron star, or black hole) accreting from a stellar companion and powered by the resulting release of gravitational energy, and their accretion processes are as rich and complex as the evolutionary histories that produce them. Beyond differences due to the type of compact object, the behavior of XRBs depends on the binary mass ratio, separation and orbital period; the mass, temperature, winds, and evolutionary stage of the donor star; and the spin and magnetic field of the compact object. Leaving aside the extensive sub-classifications \cite[for a recent review, see][]{Bahramian2022}, we note that XRBs are categorized as low- or high-mass X-ray binaries (LMXB or HMXB, depending on whether the companion star is less or greater than some threshold mass $\sim3M_\odot$) and as persistent or transient (depending on whether their accretion is continuous or episodic). X-ray binaries also exhibit a wide variety of ``accretion states" \cite{Remillard2006}, corresponding to different luminosity, spectra, and configurations of the accretion flow. Overall, they provide an excellent set of opportunities to understand the physics of accretion via X-ray spectroscopy.

If X-ray astronomy was already photon-starved (note, for example, that compared to optical CCDs, X-ray detectors are typically designed to record the details of each individual event), the push to higher spectral resolution---distributing observed counts across a much larger number of energy bins---presents a significant observational challenge. Herein lies one of the great advantages of Galactic X-ray binaries for high-resolution spectroscopy: their sheer proximity! Compared to more luminous but much more distant AGN, for example, XRBs provide many more photons per second. Furthermore, their much shorter variability timescales \cite{McHardy2006} make it possible to track major changes in their accretion flows on human timescales: a career, a PhD thesis, a summer internship, a 1 ks snapshot. At the high luminosities typical of Galactic transients, these facts facilitate high-resolution spectral  variability studies at scales that simply aren't currently feasible in other systems \cite{Nowak2017}. Nevertheless, much of the underlying physics we discuss below, such as wind driving mechanisms (\S \ref{sec:driving}) may apply to AGN and ULX studies; we refer the interested reader to Chapters 10 and 14 of this volume by Gallo \& Miller and Pinto, respectively.

The bulk of this chapter focuses on what we have learned about accretion from high-resolution X-ray spectroscopy of black hole and neutron star X-ray binaries. Much of our discussion concerns studies of accretion disk winds and the accretion-ejection connection (Section~\ref{sec:winds}). In addition, we review what high-resolution X-ray spectra have taught us about the accretion geometry in these systems (Section~\ref{sec:geometry}), and how these allow us to study the gas and dust properties of the interstellar medium (ISM; Section~\ref{sec:ism}). Finally, we provide an outlook of what the next generation of high-resolution spectroscopic instruments may bring us (Section~\ref{sec:future}).

\section{Accretion and Ejection}\label{sec:winds}

\begin{svgraybox}
The most sensitive indicators of mass loss from cosmic sources are the resonance absorption lines of abundant ions \cite{Brandt2000}.\\

\noindent \dots one would expect the mass flow rate and wind density to be strong functions of the luminosity, such that the actual structure of the wind (i.e., density and size) can change as a function of the luminosity \cite{Lee2002}.
\end{svgraybox}

One particularly fruitful avenue for study of accretion physics has come from work on the ``disk-jet connection," i.e., the relationship between accretion states, accretion disks, and relativistic jets. Black hole transients, for example, typically follow a well-established pattern in  their outbursts. They leave quiescence in radiatively-inefficient, spectrally-hard states with strong Comptonization and optically-thick jets but transition to more thermal states as they rise (and then fall again) in luminosity; transient relativistic ejections may be observed at this state transition \cite{Esin1997,Gallo2003,Corbel2003,Fender2004,Fender2004b, Fender2009}. Similar outburst patterns have been observed in neutron star LMXBs and accreting white dwarfs (cataclysmic variables or CVs) \cite{Kording2008}.

\begin{figure}[t]
\centerline{\includegraphics[width=0.9\textwidth]{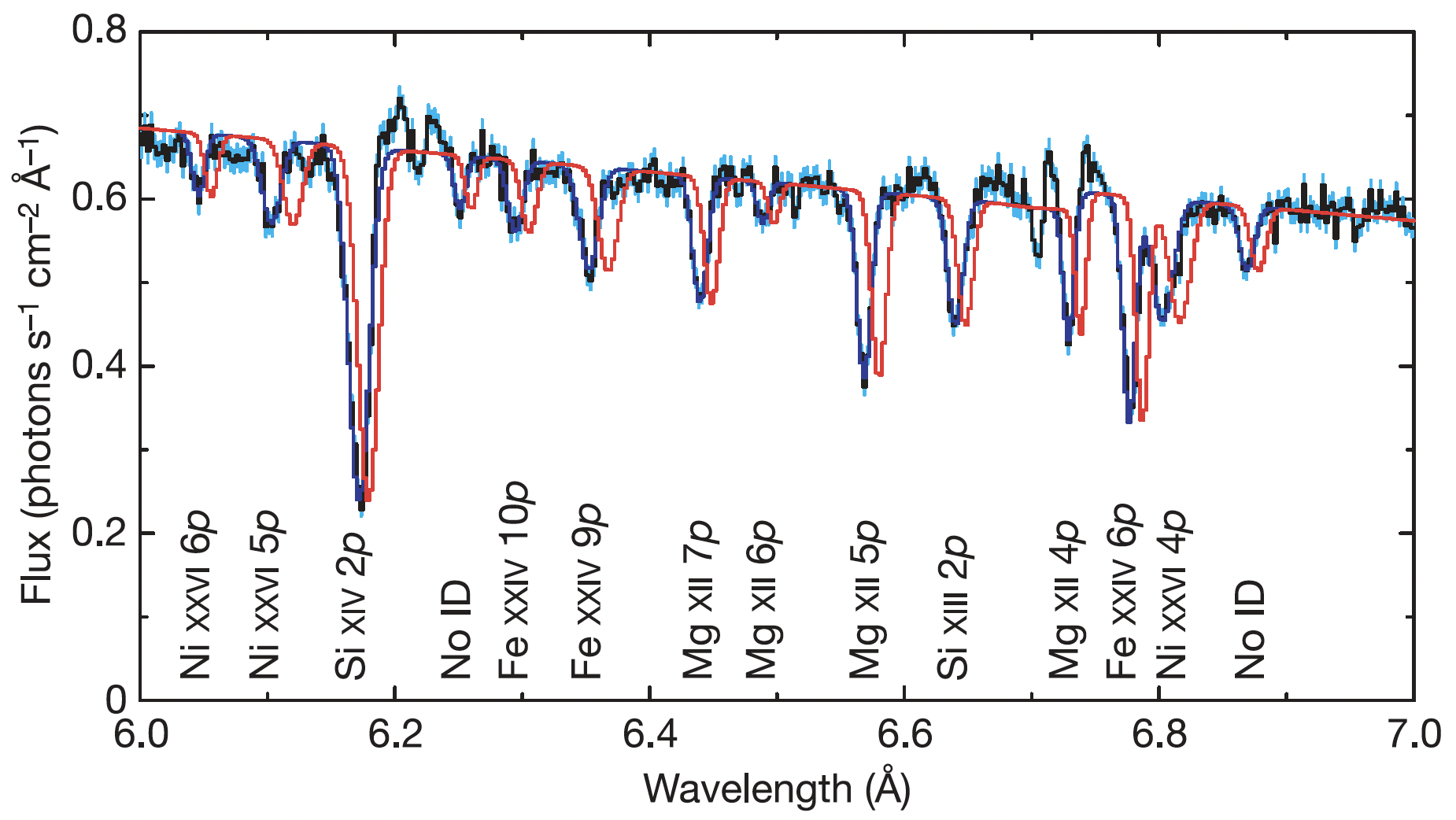}}
\caption{Blueshifted absorption lines from an accretion disk wind in the black hole LMXB GRO J1655--40 from \cite{Miller2006b}. The black curve with blue error bars is the \chandra HETGS spectrum, while the red curve is the best fit model adjusted to zero Doppler shift. High-resolution spectra like these allow sensitive diagnostics of the dynamics, temperature, ionization, and even density of absorbing gas.}
\label{fig:blueshift}     
\end{figure}

\begin{figure}[t]
\centerline{\includegraphics[width=0.9\textwidth]{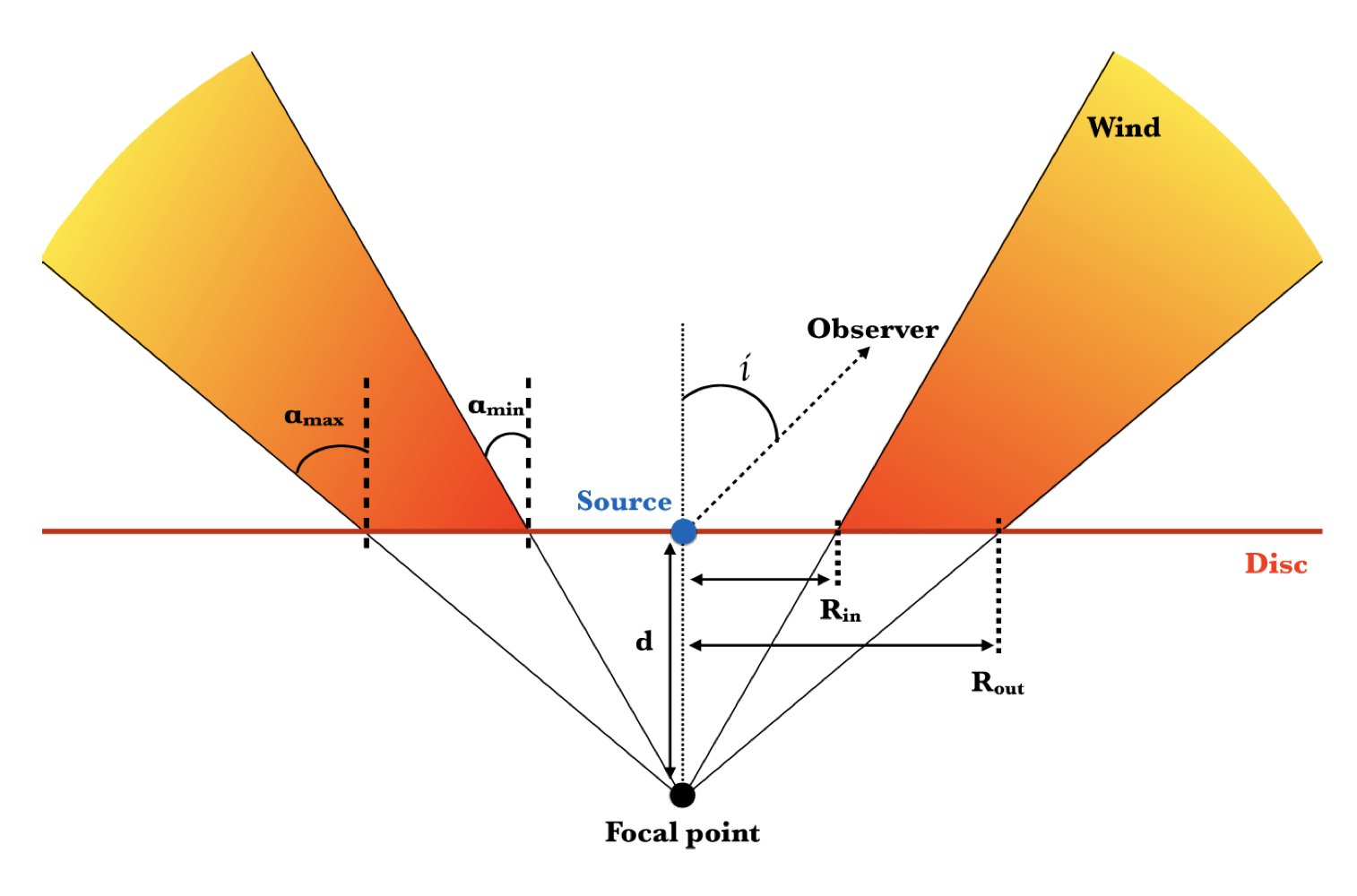}}
\caption{Cartoon of an accretion disk wind from \cite{Tomaru2018}. The wind is largely equatorial, with the densest regions and largest equivalent widths observed at high inclination, near the plane of the disk. This geometry is consistent with observational constraints from \cite{Sidoli2001,Boirin2003,DiazTrigo2006,Ponti2012} and is similar to models often used for disk winds in CVs \cite{Knigge1995,Knigge1997,Matthews2015}.}
\label{fig:cartoon} 
\end{figure}

Perhaps the most significant contribution of high-resolution X-ray spectra in this context is the unambiguous discovery of accretion disk winds in black hole outbursts. CCD spectra had revealed absorption lines in some bright X-ray sources \cite[e.g.,][]{Ebisawa1997,Kotani1997}, but without the resolution to measure Doppler shifts, it was not possible to determine whether those lines represented extended atmospheres or outflows. With grating spectra provided by \chandra \cite{Canizares2005,Brinkman2000} and \xmm \cite{denHerder2001}, it became clear that they were produced in gas flowing away from the compact object (at least in many cases, but see Section \ref{sec:geometry}). These outflows typically have blueshifts of a few hundred to a few thousand km s$^{-1}$ (Figure \ref{fig:blueshift}; see also Table \ref{tab:winds}), which can complicate the determination of an accretion disk origin when the companion star is a strong source of stellar winds. Therefore, most of our observational knowledge about disk winds in XRBs (and hence our focus here) comes from LMXBs, since their companion stars do not exhibit strong winds. Disk winds are often envisioned as biconical, roughly equatorial outflows (see Figure \ref{fig:cartoon}). They are also typically highly ionized, such that hydrogen- and helium-like iron tend to dominate the spectra with strong lines at $\sim7$ keV and 6.7 keV, respectively.

It was already evident from simulations of accretion disk winds \cite[e.g.][]{Begelman1983} that photoionized accretion disk winds would play a significant role in both the physics of accretion onto compact objects and efforts to understand them, and this prediction was quickly confirmed with early high-resolution X-ray observations. For instance, in the 2000 discovery of P-Cygni lines in Circinus X-1, \cite{Brandt2000} noted the likely connection to the high luminosity of the source. Extending this suggestion, in 2002 \cite{Lee2002} estimated the mass loss rate in the accretion disk wind in GRS 1915+105, finding it to be comparable to the accretion rate onto the black hole \cite[see also][]{Higginbottom2017}. In the two decades since, high-resolution observations of X-ray binaries have significantly enhanced our understanding of the connections between accretion and ejection processes, and in 2016 \cite{Fender2016} argued that winds could eject more than half the accreted matter during black hole outbursts\footnote{\cite{Kosec2020} found a 70\% mass loss rate in the neutron star LMXB Her X-1.}. The details tend to vary somewhat, however, depending on the system properties, particularly the nature of the compact object and whether the source is persistent or transient\footnote{Of course, whether the compact object is a black hole, neutron star, or white dwarf is related to whether it is persistent or transient via a combination of accretion physics and stellar and binary evolution.}. Here we focus mostly on black holes and neutron stars, but again many of the physical principles apply to CVs as well.

As this text is concerned primarily with X-ray spectroscopy, we also focus on high-energy observations. However, there is a long and growing list of optical, infrared, and UV observations of winds in XRBs \cite[e.g.,][]{Bandyopadhyay1997,Knigge1997b,Rahoui2014,Munoz-Darias2018,Jimenez-Ibarra2019,Sanchez-Sierras2020,Cuneo2020,Munoz-Darias2020,CastroSegura2022,MataSanchez2022,Panizo-Espinar2022} that clearly indicate that X-rays provide an incomplete picture of ionized outflows from compact objects. \cite{Munoz-Darias2022} have recently argued, for example, that accretion disk winds are likely multiphase outflows, with hot and cold gas coexisting. As next-generation X-ray missions offer higher spectral resolution and sensitivity, connecting the dots between different ouflow phases and filling in the details of the behavior of winds in black hole outbursts should be a top priority (see \S\ref{sec:future}).

\subsection{Black Holes}
\label{sec:bhwinds}

Following the early detections described above (a more detailed census is given in Table \ref{tab:winds}), studies of accretion disk winds around stellar mass black holes have largely followed two parallel but closely-related tracks: (1) quantifying the time-dependent mass loss rate \mdotw in these ionized winds and their effect on the accretion flow and the accretion state, and (2) determining the driving or launching mechanism. As detailed below, our ability to measure \mdotw is ultimately limited by our knowledge of the geometry and structure of these outflows: the launch radius and (number) density profile $n(R)$ are essential for robust estimates of mass loss rates. But because these same parameters are also needed to infer the dominant physical processes that launched any particular wind, it is not possible to fully disentangle questions of the origin of disk winds from questions of their role in systems with accreting black holes. 

\subsubsection{Photoionization}
\label{sec:ionization}
In either case, we need to be able to infer the ionization state of outflowing gas from a series of line features in a high-resolution spectrum. In X-ray observations of stellar mass black holes, these are most commonly absorption lines, but emission lines are also possible depending on the orientation and geometry of the source and the wind \cite[e.g.,][]{King2015,Sanchez-Sierras2020}. Here we present a brief overview of the physics of photoionization. For a more detailed description of astrophysical plasmas, we refer the reader to Chapter 8 of this volume by Kallman.

Consider a shell of gas of uniform electron number density $n$ at a distance $R$ from a point source with ionizing luminosity $L$; let the shell have thickness $\Delta R$ and therefore equivalent hydrogen column density \begin{equation}N_{\rm H}=n\Delta R.\end{equation} To see why a full accounting of the ionization state of the gas is necessary, imagine that this shell is ionized to the point that it produces only a single detectable line (e.g., the Fe\,{\sc xxvi} Lyman-$\alpha$ line\footnote{The Ly$\alpha$ line is a doublet, but this does not change the overall argument.}: the transition from $n=1\rightarrow n=2$ in hydrogen-like iron). Following \cite{Draine2011}, the line center optical depth is given by \begin{equation}\tau_0=\frac{\sqrt{\pi}e^2}{m_e c^2}\frac{N_\ell f_{\ell u}\lambda_{\ell u}}{b},\end{equation} where $e$ and $m_e$ are the charge and mass of the electron, $c$ is the speed of light, $N_\ell$ is the column density of Fe atoms in the lower level ($\ell$, here $n=1$), $\lambda_{\ell u}$ and $f_{\ell u}$ are the wavelength and the oscillator strength for the transition from $\ell$ to the upper level $u$ ($n=2)$, and $b=\sqrt{2}\sigma_V$ is the Doppler parameter in km s$^{-1}$. Note that this equation neglects stimulated emission.

The important detail is that the depth of the line (and its equivalent width) depends only on the column density of Fe atoms in the $\ell$ level, not on the equivalent hydrogen column density. This is related only indirectly, via \cite{Kallman2009}: \begin{equation}
    N_\ell = N_{\rm H}A_jx_{ij},
\end{equation} where $A_j$ is the elemental abundance and $x_{ij}$ is the ion fraction (i.e., the fraction of Fe atoms in the hydrogen-like state). In short, to infer the total column density of material along the line of sight, we need detailed information about electron populations in all the relevant ionization states of its abundant elements.

Per \cite{Tarter1969}, this ionization equilibrium is especially sensitive to the \textit{ionization parameter} \begin{equation}
    \xi=\frac{L}{nR^2}\label{eq:xi}
\end{equation} as well as the spectrum of the ionizing radiation field. To model a series of ionized lines, we supply a continuum model to photoionization codes like {\sc xstar} \cite{Kallman1982,Kallman2001,Bautista2001}, {\sc cloudy} \cite{Ferland1998}, or {\sc pion} \cite{Kaastra1996,mehdipour2015}. These calculate the physical conditions in gas in photoionization equilibrium, accounting for Compton heating and cooling, photoionization and recombination processes, and so on, making it possible to fit line spectra for $\xi$ and $N_{\rm H}$ directly.

If our shell subtends solid angle $\Omega$ and has velocity $v$, the associated mass loss rate will be \cite{Lee2002,Reynolds2012} \begin{equation}\dot{M}_w=4\pi m_p v nR^2\frac{\Omega}{4\pi},\label{eq:mdotn}\end{equation}  where $m_p$ is the mass of the proton. It can be difficult to constrain $n$ and $R$ directly from observations (see \ref{sec:driving}), but following \cite{Lee2002} it is possible to use Equation \ref{eq:xi} to rewrite Equation \ref{eq:mdotn} in terms of observables  as \begin{equation}\dot{M}_w=4\pi m_p v\frac{L}{\xi}\frac{\Omega}{4\pi}.\end{equation}

\subsubsection{Inferring Evolving Winds in Outbursts}
\label{sec:mdot}

The first X-ray observations of accretion disk winds in X-ray binaries also brought the first evidence of their variability. \cite{Schulz2002a} discovered P-Cygni profiles around the neutron star Cir X-1 that effectively shifted between pure absorption and pure emission over the course of their \textit{Chandra} HETGS observations. While some changes in the appearance of P-Cygni profiles can be explained by varying the illumination or orientation of the outflow, \cite{Schulz2002a} found that changes in the structure or density of the wind were required to explain the line variability in Cir X-1.

\cite{Miller2006a} reported a similar result from the 2003 outburst of the black hole LMXB H1743--322. A series of \chandra HETGS observations revealed Fe\,{\sc xxv} and Fe\,{\sc xxvi} absorption lines that varied from one pointing to the next, likely from an accretion disk wind with a mass loss rate roughly $\sim5\%$ of the Eddington rate. Interestingly, within all four observations, the lines appeared to be stronger at lower X-ray flux. One observation in particular, however, showed even stronger variability centered around a $\sim300$ s oscillation in the X-ray lightcurve. Flux-resolved grating spectroscopy of this oscillation revealed changes not only in the depth of the lines but also in their ratio: the depth of the Fe\,{\sc xxv} line increased much more than the depth of the Fe\,{\sc xxvi} line at low flux. As noted by \cite{Miller2006a}, this could be explained by a decrease in the ionization of the gas at low flux, but it could also be indicative of changes in the density structure or geometry of the absorption, e.g., an inhomogeneous wind. \cite{Neilsen2011} pushed similar techniques further, measuring wind variability in GRS 1915+105 on timescales of seconds but also finding evidence of fast structural variability.

As the catalog of accretion disk wind sources grew, it  became clear that these outflows exhibited long-term variability in addition to the faster changes described above. For instance, \cite{Miller2008} noted that absorption lines from accretion disk winds were stronger during softer (e.g., disk-dominated) states. This pattern was reflected in observations of H1743--322 \cite{Miller2006a} and GRO J1655--40 \cite{Miller2006b} as well as archival observations of GRS 1915+105. These latter observations were published in detail in \cite{Neilsen2009b}, who characterized the state dependence as broad emission lines in harder states (consistent with reflection) and narrow absorption lines in softer states.

\begin{figure}[t]
\centerline{\includegraphics[width=0.85\textwidth,angle=270,clip=true,trim=10 10 10 0]{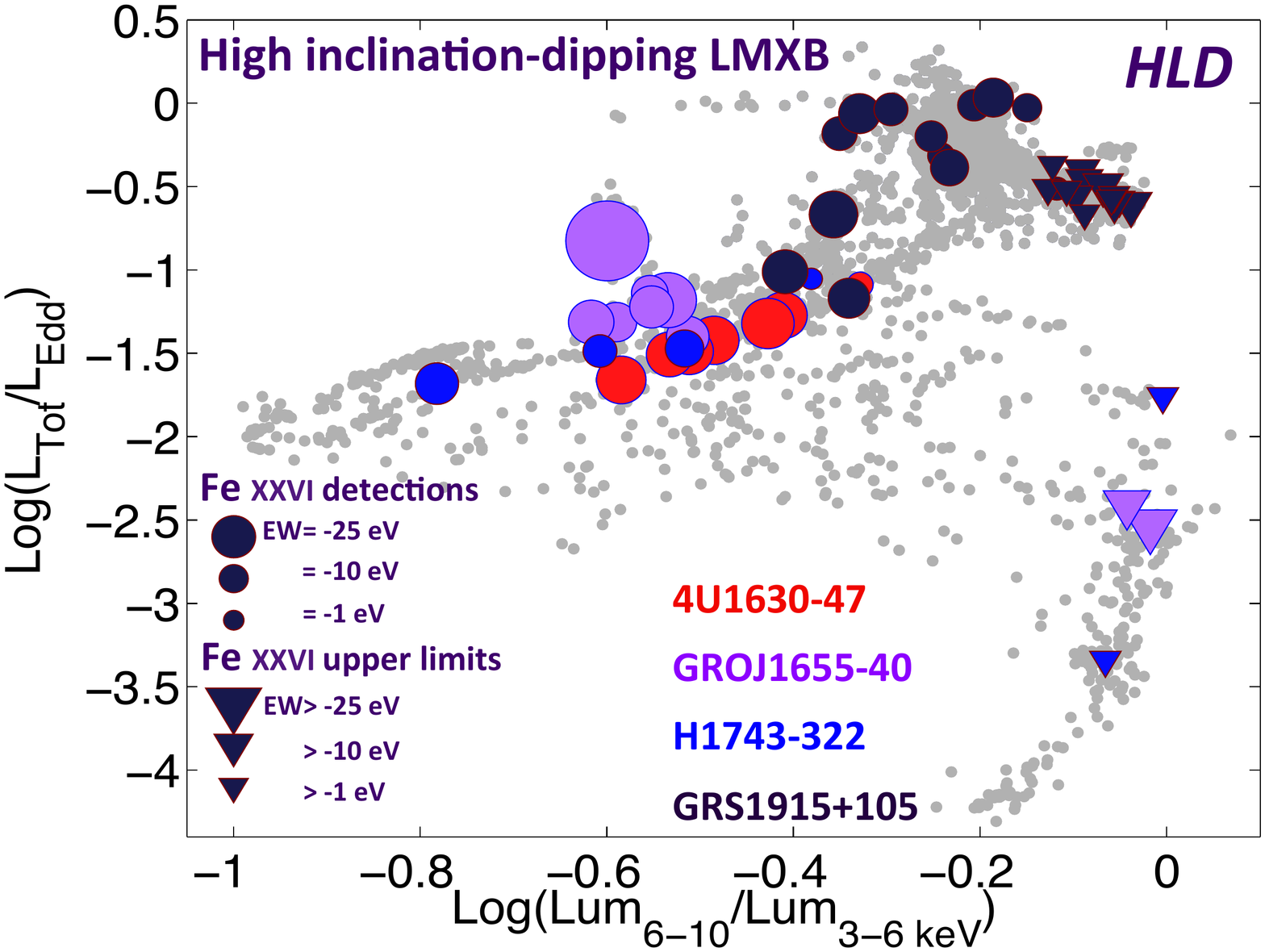}}
\caption{Observations of accretion disk winds in high inclination black hole X-ray binaries from \cite{Ponti2012}. Background gray points show the hardness-intensity trajectories of outbursts. Circles show wind detections, with symbol size scaling with absorption line equivalent width. Triangles indicate non-detections, again with symbol size proportional to the upper limit on any line equivalent width.
\label{fig:ponti}}      
\end{figure}

If accretion disk winds vary by state, it is natural to wonder whether there is a connection between winds and radio jets \cite[given the well-characterized link between jets and X-ray hard states, e.g.,][]{Fender1999,Fender2004,Fender2009}. This connection must be complex, because both \cite{Lee2002} and \cite{Miller2006a} reported non-zero radio flux coincident with detections of absorption lines. \cite{Miller2008} suggested that an anticorrelation between winds and jets might relate to changes in the wind density and its effect on the  magnetic field configuration in the inner disk. In contrast, \cite{Neilsen2009b} interpreted the state dependence of accretion disk winds as evidence of a material tradeoff between winds and jets, in which mass loss in winds from the outer accretion disk (see \S\ref{sec:driving}) effectively reduces the amount of matter fed into the jet launched from the inner disk.

The picture of this anticorrelation is substantially enhanced with additional data. \cite{Ponti2012} performed a systematic search for accretion disk wind absorption lines (specifically Fe\,{\sc xxv} and Fe\,{\sc xxvi}) in X-ray observations of black hole X-ray binaries. In systems seen at high inclination, they found that iron absorption lines were preferentially detected in softer states, generally at higher luminosity (see Figure \ref{fig:ponti}), matching the early trend described in \cite{Miller2008,Neilsen2009b}. Interestingly, they detected no significant  hot wind absorption in any system at low inclination, which is consistent with the idea of disk winds as equatorial outflows \cite[e.g.,][]{Boirin2003,DiazTrigo2006}. The only apparent counterexample is a warm absorber in GX 339-4 \cite{Miller2004}, but this is apparent only in Ne, O, and Mg below 1.5 keV and its origin is not entirely clear \cite[but see][]{Luo2014}. See Table \ref{tab:winds} for an updated census of winds in BH XRBs.

While \cite{Ponti2012} has often been cited as evidence that X-ray absorbing accretion disk winds are never found in X-ray hard states or that jets and such winds cannot coexist, the historical record described above requires a more nuanced view. \cite{Miller2006a} found strong radio activity coincident with X-ray absorption lines in H1743--322, and \cite{Homan2016} reported similar results at high luminosity in GRS 1915+105, V404 Cyg, GX 13+1, Sco X-1, and Cir X-1. Thus it seems clear that any interaction between winds and jets---whether mediated by mass loss or changes in the magnetic field configuration---must at the very least be gradual, so that both outflows can be observed simultaneously.

Regardless of the precise timescale for the transition to wind-dominated states, Figure \ref{fig:ponti} \cite[taken from][]{Ponti2012} suggested that there is a real trend in the observability of absorption lines in black hole outbursts. The discovery of X-ray absorption lines in \chandra HETGS observations of 4U 1630--47 during its 2012-2013 outburst lent additional credibility to the idea that winds can be reliably detected in soft states of high-inclination systems \cite{Neilsen2013,Neilsen2014}; see also \cite{Miller2015,Trueba2019}. But if that fact is generally accepted, the apparent absence of winds in hard states is still not fully understood, partly because several processes can affect their observability, including (1) overionization, (2) thermodynamics, and (3) time evolution.

\textbf{Overionization:} Implicit in our discussion of ionization balance and the definition of the ionization parameter $\xi$ (Equation \ref{eq:xi}) is the shape of the ionizing spectrum. Consider two identical shells of gas equidistant from X-ray sources with luminosity $L;$ let one be illuminated by a blackbody spectrum and the other by a power law spectrum. The gas shells will have the same ionization parameter but very different ion fractions. These depend on the integral of the photon rate spectrum over the ionization cross section $\sigma_i(E)$ of the relevant charge state, i.e.: \begin{equation}\Phi_i=\int_{\chi_i}^\infty \frac{L_E}{E}\sigma_i(E)dE,\label{eq:phi}\end{equation} where $E$ is energy, $L_E$ is the monochromatic luminosity and $\chi_i$ is the ionization threshold for charge state $i$. For a given luminosity,  a harder/shallower spectrum will generally ionize each charge state shell more effectively. For ionization parameters of interest here, $\log(\xi)\approx4-5$, a harder ionizing spectrum will typically lead to weaker iron absorption lines.

The idea behind ``overionization" as an explanation for the apparent absence of winds in spectrally hard states \cite[e.g.,][]{Neilsen2012b,Miller2012,Neilsen2013,Ponti2015,Shidatsu2019,Petrucci2021} is that winds might be present during these states but---given their harder spectra---too ionized to produce detectable absorption lines. The difficulty with overionization is that for a given luminosity, a harder spectrum may have \textbf{fewer ionizing photons} than a softer spectrum due to the factor of $E^{-1}$ in Equation \ref{eq:phi}. Moreover, from Figure \ref{fig:ponti}, much of the hard state at issue actually has lower luminosity than the soft states where winds are ubiquitous. Accordingly, one should expect a given parcel of gas to have a \textit{lower} ionization parameter during the hard state, which can compensate for the shift in the ionization balance\footnote{Given the differences in luminosity, a proper statistical analysis of the presence or absence of winds in different outburst states needs to account for the 7 keV sensitivity of each individual observation.}. Even power law spectra at higher luminosities have repeatedly proven insufficient to fully ionize disk winds present in soft states \cite{Neilsen2012b,Ponti2015}. Thus, though the details of ionization are certainly important, overionization is not likely to be the primary driver of the state dependence of hot winds in black hole X-ray binaries \cite[][reached a similar conclusion based on models of thermally driven winds]{Done2018}. However, overionization may play a more important role in optical winds \cite[e.g.,][]{Munoz-Darias2019}.

\textbf{Thermodynamics:} In order to detect Fe\,{\sc xxv} and Fe\,{\sc xxvi} absorption lines (or any other particular line series) from our shell, it needs to maintain an appropriate temperature and ionization parameter over the course of an observation. This is not particularly likely if the shell is not in a stable thermal and ionization equilibrium. The question is then whether---given all the relevant heating, cooling, and ionization processes---such an equilibrium is more or less likely during hard states than soft states. This has been studied extensively, and while it is still necessary to account for the actual radiation field in any specific observation, the broad conclusion is fairly robust that gas capable of producing hot iron absorption lines is probably thermally unstable during spectrally hard states \cite[e.g.,][]{Chakravorty2013,Chakravorty2016,Bianchi2017,Petrucci2021,Chakravorty2023}. 

\begin{figure}[t]
\centerline{\includegraphics[width=\textwidth]{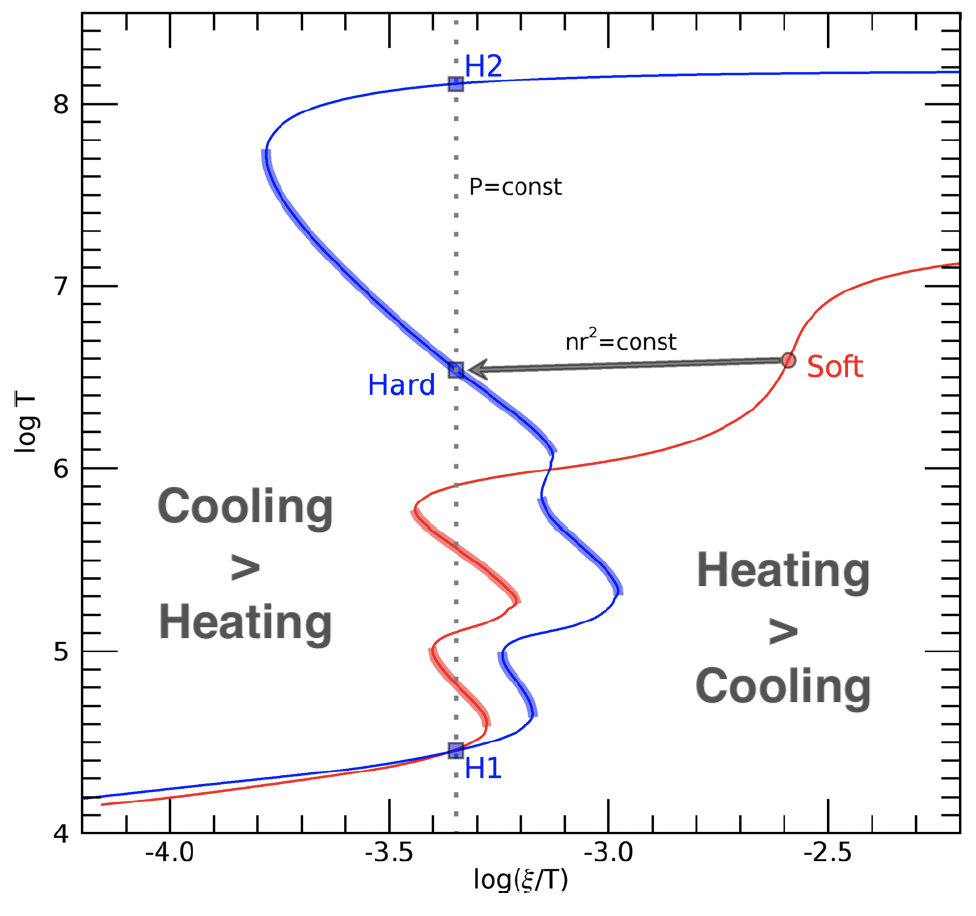}}
\caption{Stability curves for a disk atmosphere in AX J1745.6--2901, adapted from \cite{Bianchi2017}. The red curve represents thermal equilibrium for gas illuminated by the soft spectrum observed by \xmm and \nustar \cite{Ponti2015}. The best-fit parameters for the absorber are shown as a red circle, which lies on the stability curve where the slope is positive. The blue curve, on the other hand, corresponds to thermal equilibrium for gas illuminated by the hard state. Adjusting for the luminosity of the hard state, the same disk atmosphere would be thermodynamically unstable (see the point labeled ``Hard," where the slope is negative). Gas under these conditions would rapidly evolve to one of the points labeled ``H1" or ``H2."
\label{fig:stability}} 
\end{figure}

Typically thermal stability is diagnosed by inspecting \textit{stability curves}: the set of points corresponding to thermal equilibrium in a plot of $T$ vs $\xi/T$ \cite{Krolik1981}. The stability curve partitions the parameter space: to the left of the curve, the cooling rate exceeds the heating rate, while the reverse is true to the right of the curve. Our shell will only be in a thermodynamically stable equilibrium where the slope of the stability curve is positive (as this is where thermally perturbed gas will return to equilibrium)\footnote{As discussed by \cite[][see also \citenum{Goncalves2007}]{Proga2022}, this analysis is appropriate for steady illumination of an optically thin medium; alternative stability curves give similar results in the case of time dependence or non-negligible attenuation.}. Consider the example shown in Figure \ref{fig:stability}. \cite{Ponti2015} observed the neutron star AX  J1745.6--2901 on numerous occasions with \xmm and \nustar, finding absorption lines---likely from an accretion disk atmosphere---during soft states and none during hard states. \cite{Bianchi2017} used their constructed hard and soft SEDs to compute stability curves for this absorbing gas. 

As shown, they found that the absorber lay on the stable portion of the stability curve during the soft state, but that the same gas would be thermodynamically unstable during the hard state. Thus, during the hard state, the same gas would rapidly evolve to a much higher or much lower temperature and ionization than observed in the soft state. As hard spectra often render $\log(\xi)=3-5$ gas thermodynamically unstable, this is a very compelling explanation\footnote{One important complicating factor noted by \cite{Done2018} is that state changes in the broadband spectrum lead to changes in the Compton temperature (see \S \ref{sec:driving}). One consequence is that thermally driven winds may be launched from different locations in the disk during hard and soft states. If the value $nR^2$ is not preserved in the wind during a state transition, then it becomes much more difficult to predict the expected temperature and ionization of the wind in the hard state. Robust conclusions about overionization thermal instability may therefore depend on the inferred driving mechanism.} of X-ray visibility trends for accretion disk winds. It is worth noting that \cite{Higginbottom2020} performed radiation hydrodynamic simulations of disk winds, including a self-consistent ionization and frequency structure, including radiative driving (see below) and thermal instabilities but still found that hot winds should be detectable in hard states at a high inclination ($i>75^\circ$). The source of this discrepancy is not clear, since \cite{Miller2012} placed tight upper limits on absorption line equivalent widths in hard states of H1743--322.

Nevertheless, thermodynamic instabilities also offer an appealing way to account for the fact that winds are rarely detected as X-ray absorption lines during hard states, but that they \textit{are} detectable in the optical and infrared during these states. For example, \cite{Munoz-Darias2016} detected P-Cygni profiles of optical emission lines coincident with radio jet emission during the 2015 outburst of V404 Cygni, and \cite{Munoz-Darias2018} found similar results during likely hard states of V4641 Sgr.  \cite{Munoz-Darias2019} observed cold optical winds throughout the rising and decaying hard states in MAXI J1820--070 (but not in the soft state) and suggested that these outflows are a common feature of accreting black hole systems. Interestingly, \cite{Sanchez-Sierras2020} found evidence for a relatively steady (i.e., state-independent) wind in NIR spectra of the same outburst. But finally, \cite{Munoz-Darias2022} have reported the simultaneous detection of X-ray and optical signatures of accretion disk winds in V404 Cyg; they argue that the data are consistent with a multiphase outflow, i.e., hot and cold gas coexisting in a single wind \cite[see also][]{Waters2021}. Therefore disappearance of X-ray absorption lines need not indicate the disappearance of the wind itself.

\textbf{Time Evolution:} In other words, the fact that accretion disk winds may be very difficult to observe in the X-ray during spectrally hard states does not mean that they are not present during these states. Indeed, \cite{Tetarenko2018} argued that massive accretion disk winds shape the lightcurves of black hole outbursts, including during the hard state. As noted above, optical and IR spectra of black hole disk winds in hard states help to reveal how the high-energy signatures of these outflows may be shaped by thermodynamic instability.

But there are also several definitive instances of state-dependent evolution of hot accretion disk winds that cannot be explained by ionization or thermodynamics. For example, two observations of an accretion disk wind in the 2005 outburst of GRO J1655--40 were inconsistent with a static wind \cite{Neilsen2012b}. In addition, \cite{Gatuzz2019} showed that overionization and thermal instabilities could not account for the disappearance of the accretion disk wind in the 2012-2013 outburst of 4U 1630--47 \cite[see][]{Neilsen2013,Neilsen2014}.

While these examples do not pertain directly to the apparent absence of winds in most spectrally hard states, they provide compelling evidence that the physical properties of accretion disk winds evolve significantly over the course of black hole outbursts. Theoretically, this does not come as a surprise: depending on the driving or launch mechanism of the wind, the mass loss rate is expected to change with accretion state (e.g., the geometry and spectrum of the accretion flow). For instance, \cite{Ponti2012} noted that if the wind is driven by irradiation or Compton heating of the outer accretion disk (i.e., ``thermal driving"), any changes in the illumination of the outer disk---such as might be caused by an increase in the disk scale height during hard states---could lead to less effective wind driving. As discussed above and below, changes in the broadband X-ray spectrum are also expected to affect thermal wind launching, introducing a time-dependent $\dot{M}_w$ \cite[see also][]{Done2018}. But magnetohydrodynamic (MHD) winds may also evolve during outbursts: \cite{Fukumura2021} inferred changes in the wind density profile in different accretion states, though the theoretical basis for those changes is not understood.\vspace{3mm}

In summary, there is strong evidence that photoionization, thermodynamics, and true physical time dependence all shape the appearance of accretion disk winds in X-ray binaries. Now we turn to the driving mechanisms, which (in addition to being important in their own right), will shed some additional light on the physical time dependence of winds.

\subsubsection{Wind Driving Mechanisms}
\label{sec:driving}

From the preceding discussion, it should be clear that efforts to interpret the results of \cite{Neilsen2009b, Ponti2012} and understand the mass loss rate in winds, we need to understand the physical processes that launch these outflows from the accretion disk in the first place. There has been a significant effort in recent years to infer these mechanisms based on observations \cite[e.g.,][]{Miller2006b,Netzer2006,Kallman2009,Neilsen2013,Neilsen2016,Shidatsu2016,Kalemci2016,Fukumura2017,Fukumura2021,Tomaru2022}. Following decades of literature, we shall consider three main driving mechanisms: (1) radiation pressure, (2) thermal driving, and (3) MHD processes. It is common in observational papers to consider these mechanisms separately, though (a) hybrid winds are allowed and often distinct from the main categories \cite{Proga2000,Proga2003,Waters2018,Done2018,Shidatsu2019,Higginbottom2020}, and (b) all three mechanisms may be operating at some level in the disk at any given moment. Here we briefly discuss each mechanism in turn and discuss the overlap between them as it arises. Efforts to determine the (dominant) driving mechanisms of winds are typically a process of elimination, so where possible, we focus on observational diagnostics of and constraints on the driving mechanism.

\textbf{Radiation Pressure:} In a radiation pressure driven wind, photons scatter off atoms or electrons in the vicinity of the compact object, accelerating the gas until it reaches some terminal speed in excess of the local escape velocity. Here the process that accelerates the wind is the transfer of momentum from the radiation field to the gas \cite{Reynolds2012}. This mechanism seems intuitively relevant for X-ray binaries, which often reach high luminosities and therefore might be expected to launch radiatively-driven winds efficiently.

The reality is rather more complicated. Briefly, radiation driving requires a source of opacity for scattering, and is therefore especially sensitive to the ionization state of the gas \cite{Vitello1988}. For a fully ionized gas, that opacity is due to electron scattering. At lower temperatures, there is significantly more opacity in atomic lines, so that resonant scattering or ``line driving" \cite[e.g.,][] {Murray1995} is much more effective at launching winds from accretion disks. This is especially the case for the UV, where the line opacity is particularly high. To be quantitative, the line force can exceed that due to electron scattering by $\lesssim2000\times$ \cite{Castor1975,Abbott1982,Gayley1995}. 

In detailed radiation hydrodynamic simulations, \cite{Proga2002} found that radiation pressure could only effectively drive winds in LMXBs when the central luminosity is within a factor of $\sim2$ of the Eddington limit\footnote{\cite{Ueda2004} argued for the effectiveness of radiation-driven winds at similar luminosities on the idea that for the purposes of gas kinetics, radiation pressure reduces the effective mass of the compact object, and that the gas is effectively unbound at high Eddington ratios.}, but that such winds are best described as thermally driven (see below) with assistance from electron scattering \cite[see also][]{Higginbottom2020}. Line driving was not effective at ionization parameters $\log(\xi)\gtrsim3$ due to the reduction in UV opacity. Thus when we encounter a wind absorption line system with a measured high ionization parameter in X-ray binaries, we typically disregard radiation pressure as a viable driving mechanism for that particular outflow. 

Practically speaking, this criterion rules out radiation pressure as the source of winds in most black hole and neutron star X-ray binaries, with a few notable exceptions. For example, \cite{Ueda2004} argued that the wind in the neutron star GX 13+1 (see below) could be explained as a radiatively-driven wind launched at high Eddington ratio, though \cite{Tomaru2020a} recently modeled \chandra HETGS 1st- and 3rd-order spectra of this source as a radiative-thermal wind. \cite{Proga2005} argued that line driving is the most plausible origin for disk winds in CVs \cite[see also][and references therein]{Pereyra2004}.

\begin{figure}[t]
\centerline{\hfill\includegraphics[width=0.45\textwidth]{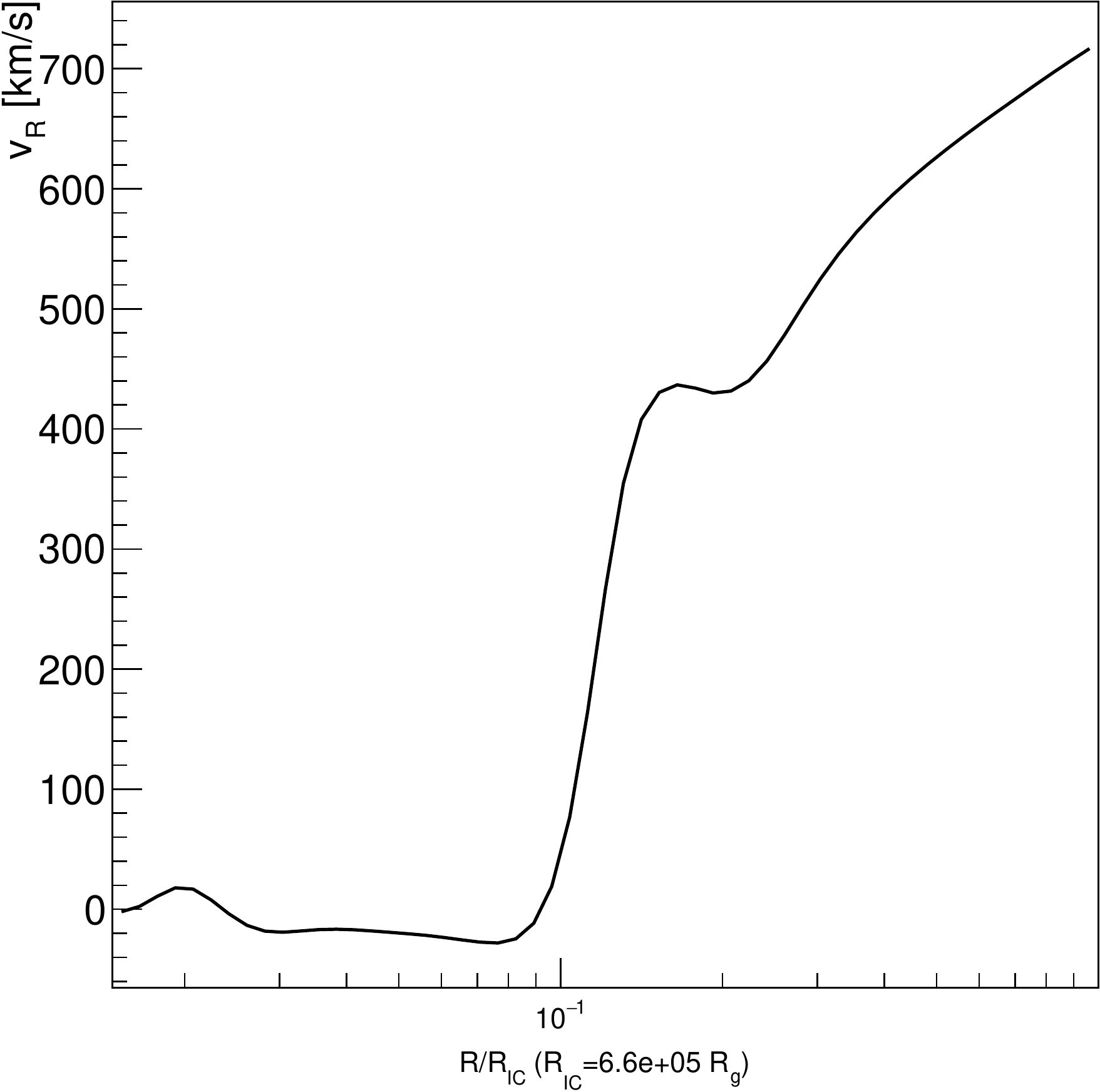}\hfill\includegraphics[width=0.45\textwidth]{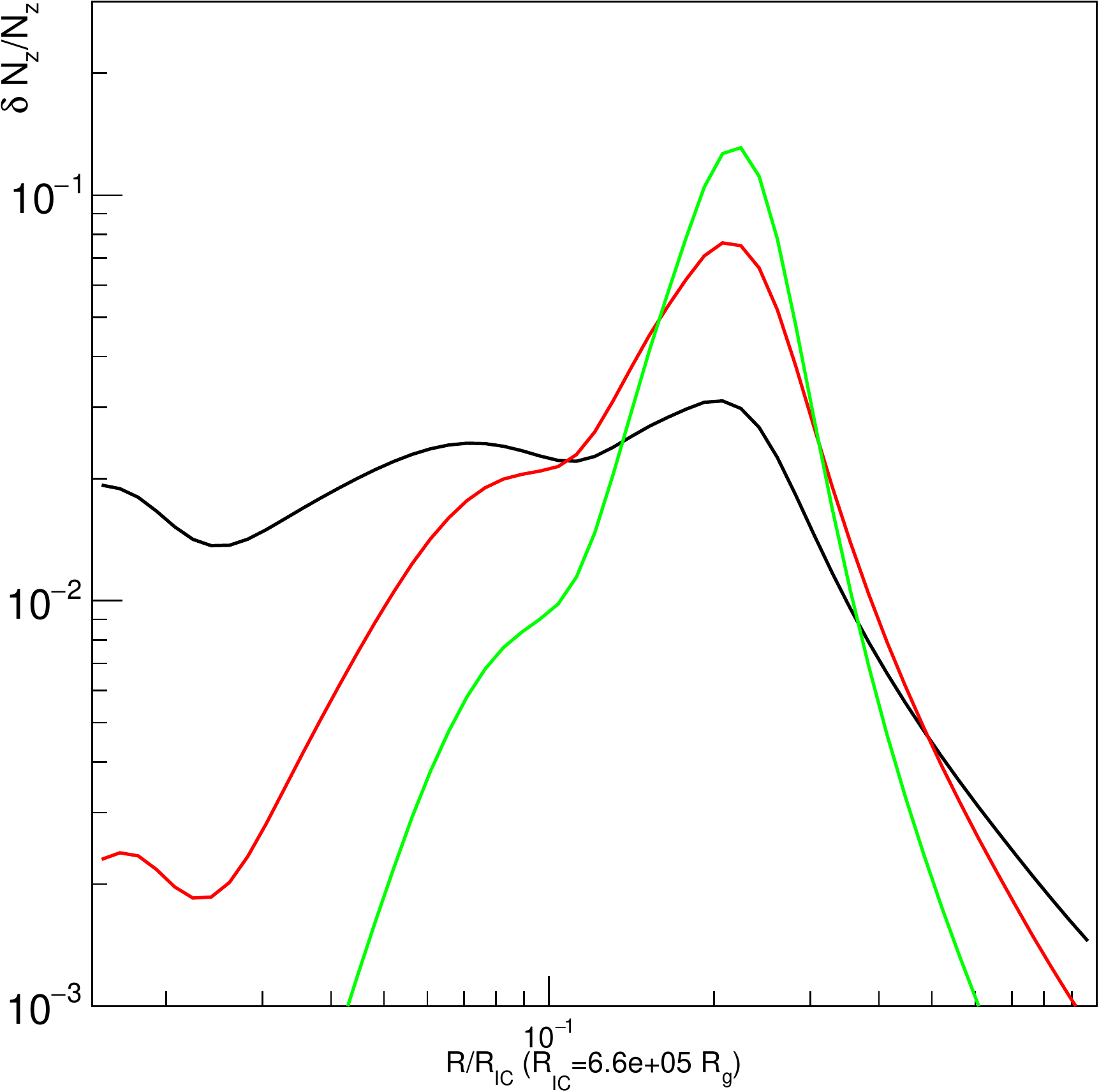}\hfill}
\caption{Calculations from 
\cite{Tomaru2020b} showing the radial velocity profile (left) and fractional ionic column densities (i.e., scaled to the total; right) for a thermal-radiative wind in H1743--322. These profiles are computed with Monte Carlo radiative transfer using a radiation hydrodynamic simulation as input. In the right panel, red and green indicate Fe\,{\sc xxvi} and Fe\,{\sc xxv}, respectively, while the black curve shows the total column density. The blueshift and ionization are strong functions of position, so comparing these models to data permits estimates of the location and turbulent velocity in the wind (see their Figure 9).
\label{fig:vprofile}} 
\end{figure}

\textbf{Thermal Driving:} If the accretion disk is sufficiently heated by X-rays from the inner accretion flow, the average thermal velocity of the irradiated disk surface can exceed the local escape velocity, especially in the outer disk where matter is less tightly bound to the central source \cite{Begelman1983}. Here we briefly follow \cite{Rahoui2010} to cover the important points. Matter in the disk can be heated to the Compton temperature \begin{equation}
T_C = \frac{1}{4k_B}\frac{\int_{0}^{\infty}EL_EdE}{\int_{0}^{\infty}L_EdE},
\end{equation} where $k_B$ is the Boltzmann constant; $T_C$ is often in the range $10^6-10^8$ K for typical XRB spectra. The characteristic size scale for a thermally-driven or \textit{Compton heated} wind is the Compton radius $R_{IC}$, where the local escape velocity equals the thermal velocity at the Compton temperature: \begin{equation}
    R_{IC}=\frac{9.8\times10^{17}}{T_C}\frac{M}{M_\odot}\,{\rm cm}, \label{eq:Rc}
\end{equation} where $M$ is the mass of the central object and $T_C$ is given in K. The Compton radius is often comparable to the size of the accretion disk, but hydrodynamic simulations \cite{Woods1996} have indicated that thermal winds can be launched from radii $R\gtrsim0.1R_{IC}$; the inclusion of radiation pressure may enable thermal-radiative winds to be launched from $\sim$anywhere in the disk at high Eddington ratio \cite{Done2018}. Finally, in order for the wind to be heated fast enough to overcome gravity, the luminosity must exceed a critical threshold \begin{equation}
    L_{\rm cr}=\frac{288L_{\rm Edd}}{\sqrt{T_C}},
\end{equation} where $L_{\rm Edd}$ is the Eddington limit and again $T_C$ is given in K. Because of the appearance of the luminosity and spectral shape in these equations, thermally-driven winds are naturally dependent on the accretion state, with a complex tradeoff between the increase in luminosity and decrease in spectral hardness (and $T_C$) as sources head from hard states to soft states \cite{Higginbottom2019}.

With the exception of observations at  a high Eddington ratio, the primary diagnostic of thermal driving is whether or not the X-ray absorber can be located outside the Compton radius (or the equivalent $0.1-0.2R_{IC}$). If so, then thermal driving could be the dominant mechanism responsible for the wind. There are several ways to estimate the location of a wind. \cite{Miller2006b,Miller2008,Kallman2009} used density-sensitive line features to measure the density of a wind in GRO J1655--40; if $L,$ $n$, and $\xi$ are known, one can solve Equation \ref{eq:xi} to determine $R$, though this formulation implicitly assumes a point source emitter \cite{Neilsen2016}. \cite{Neilsen2009a} reported a similarly complex absorber in GRS 1915+105; \cite{Ueda2009} found a large photoionization radius ($\gtrsim10^5r_g$), consistent with a radiative/thermal wind. \cite{King2015} found strong emission lines and P-Cygni profiles at high luminosity in the 2015 outburst of V404 Cyg; ionization analysis implied a large radius and a wind driven by radiation or irradiation.  Some variability arguments \cite[e.g.,][]{Neilsen2011,Neilsen2012a} have  enabled estimates of the location of X-ray absorbing gas and led to the inference of thermally driven winds. \cite{Neilsen2009b} assumed that the blueshift of absorption lines in GRS 1915+105 represented the local escape velocity, which placed the absorption lines in the outer accretion disk. Similarly, the velocity profile of observed absorption lines may be compared to predictions of models based on different launch mechanisms \cite[Figure \ref{fig:vprofile};][]{Ueda2004,Tomaru2020a,Tomaru2020b}.

Studies of thermally-driven winds have benefited enormously from recent advances in simulations, which have facilitated direct comparisons between models and observations \cite{Luketic2010,Higginbottom2015,Higginbottom2018,Shidatsu2019,Tomaru2020a,Tomaru2020b,Higginbottom2019,Higginbottom2020}. As a result, we also have a deeper understanding of the connection between atomic processes and the bulk behavior of winds. For example, \cite{Higginbottom2015} pointed out that the thermodynamic instability discussed in Section \ref{sec:mdot} is where winds are accelerated most efficiently: when plasma becomes thermally unstable, its temperature  rapidly approaches the Compton temperature \cite[see also][]{Begelman1983}. The sudden increase in the gas temperature provides a significant boost to thermal wind launching. \cite{Dyda2017} found that winds may undergo several stages of acceleration if the irradiating flux is sufficient to drive the thermal equilibrium curve through multiple steep or unstable branches. \cite{Dubus2019} demonstrated that winds themselves may represent a significant source of disk irradiation if they scatter photons down onto the disk \cite[see also][]{Kimura2019}. In short, cutting edge theoretical and observational studies of thermal winds continue to reveal the rich relationships between accretion and ejection processes in X-ray binaries.

\textbf{MHD Winds:} The same can be said for studies of winds driven by magnetohydrodynamic processes. The virtue of this driving mechanism was put concisely by \cite{Proga2003}: ``one of the reasons for favoring magnetic fields as an explanation for mass outflows from accretion disks is the fact that magnetic fields are very likely crucial for the existence of all accretion disks." One recurring theme in studies of disk winds is the connection to jets, not only owing to the related state dependence of these outflows (see above), but also because jets are fundamentally magnetic processes. Could winds and jets be separated only be changes in the magnetic field configuration? The Blandford-Payne mechanism \cite[][see below]{Blandford1982} is often discussed both as the origin of magnetocentrifugal winds \cite{Reynolds2012} and as a source of jets \cite[e.g.,][]{Neilsen2014}. In this context, \cite{Miller2006a,Miller2008,Miller2012} commented on possible magnetic connections between winds and jets. In a systematic study of AGN and XRBs, \cite{King2013} found similar wind-luminosity and jet-luminosity scaling relations and suggested a common launching mechanism.

Magnetically-driven winds can be launched from the accretion disk by several processes, one of the best known being the Blandford-Payne mechanism \cite{Blandford1982}. This mechanism is also known as a magnetocentrifugal wind, where matter from the rotating disk is flung along poloidal magnetic field lines \cite{Pelletier1992,Proga2003,Pudritz2007}. As noted by \cite{Reynolds2012}, the terminal velocity of these winds is  rather large, typically 2-3 times the escape velocity at the launch radius and 3-5 times the escape velocity at the outer edge of the acceleration region. One alternative is magnetic pressure: when the differential rotation of the accretion disk generates a large toroidal magnetic field, the gradient of the toroidal field can drive a wind off the disk \cite[][and references therein]{Proga2003}. \cite{Proga2003} performed a set of simulations to determine how magnetic fields affect radiation driven winds. They found that for sufficiently large magnetic fields, a dense, slow, pressure-driven wind dominates the line-driven component. Interestingly, \cite{Waters2018} found that including strong poloidal fields in simulations of thermal winds actually suppresses the thermal driving mechanism.

\begin{figure}[t]
\centerline{\includegraphics[width=\textwidth]{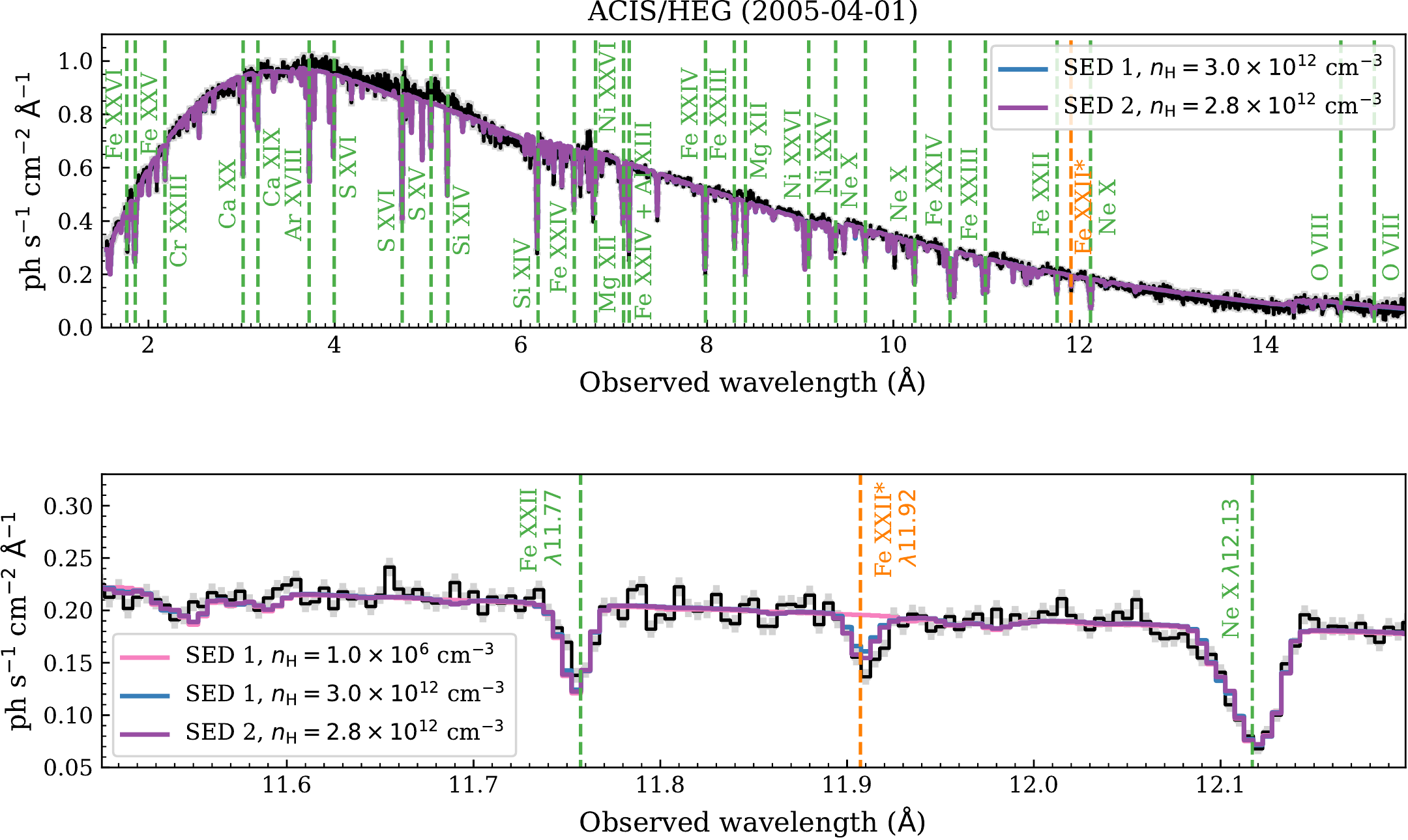}}
\caption{Spectral analysis of a rich absorption line spectrum in GRO J1655--40 from 
\cite{Tomaru2022}. The top panel shows the full absorption line spectrum including two spectral models with low density $n=\sim3\times10^{12}$ cm$^{-3}$.  The bottom panel shows a close-up of the Fe\,{\sc xxii} lines at 11.77\,\AA\, and 11.92\,\AA.
\label{fig:metastable}}   
\end{figure}

Observationally, the first claim of a magnetically-driven wind in an XRB based on high-resolution spectra was made by \cite{Miller2006b}, based on a complex absorption line spectrum in GRO J1655--40 observed by \chandra during a 2005 outburst. As noted above, \cite{Miller2006b} used density-sensitive absorption line features to infer the gas density in the wind (see Figure \ref{fig:metastable}). One constraint in particular came from the detection of Fe\,{\sc xxii} absorption lines at 11.77\,\AA\, and 11.92\,\AA\, \cite{Miller2008}, which represent transitions from the ground state $2s^22p_{1/2}$ and the metastable $2s^22p_{3/2}$ state into the $2s^23d_{3/2}$ and $2s^23d_{5/2}$ excited states, respectively. According to \cite{Mauche2003}, the ratio of these two lines is sensitive to the density in the range $n=10^{12}-10^{15}$ cm$^{-3}$ because collisional excitation can enhance the population of the $2s^22p_{3/2}$ state (leading to a stronger 11.92\,\AA\, line). By comparing the observed line ratio to the predicted density dependence, \cite{Miller2008} inferred a density $n=10^{13.7}$ cm$^{-3}$; \cite{Kallman2009} noted several other transitions that also suggest high density. \cite{Miller2014a} found a similar result in MAXI J1305--704.

Via photoionization modeling \cite{Miller2006b,Miller2008} concluded that the absorber in GRO J1655--40 lay at $R\gtrsim10^9$ cm from the black hole. Since the source appeared to be at only a few percent of the Eddington limit and the wind was highly ionized, such a small radius seemed to rule out both thermal driving and radiation pressure, leaving magnetic processes as the only apparently viable origin for the wind\footnote{\cite{Neilsen2012b} explored the time dependence of the wind in GRO J1655--40 and found that it likely had at least one thermally-driven component earlier in the outburst.}.  The low velocity of the wind \cite[$\sim375$ km s$^{-1}$;][]{Kallman2009} in GRO J1655--40 is better matched to a wind launched by magnetic pressure than rotation, since the escape velocity at the apparent radius is 15-30$\times$ the observed blueshift; \cite{Neilsen2013} also argued that the data are incompatible with a magnetocentrifugal wind based on constraints from \cite{Reynolds2012}.
The identification of this outflow as a magnetically-driven wind has inspired significant debate since 2006. Here we will discuss some additional results on MHD wind models and their applications to spectra of black hole X-ray binaries, then return to the origin of the wind in of GRO J1655--40.

\begin{figure}[t]
\centerline{\includegraphics[width=0.7\textwidth,angle=270,clip=true,trim=0 30 0 0]{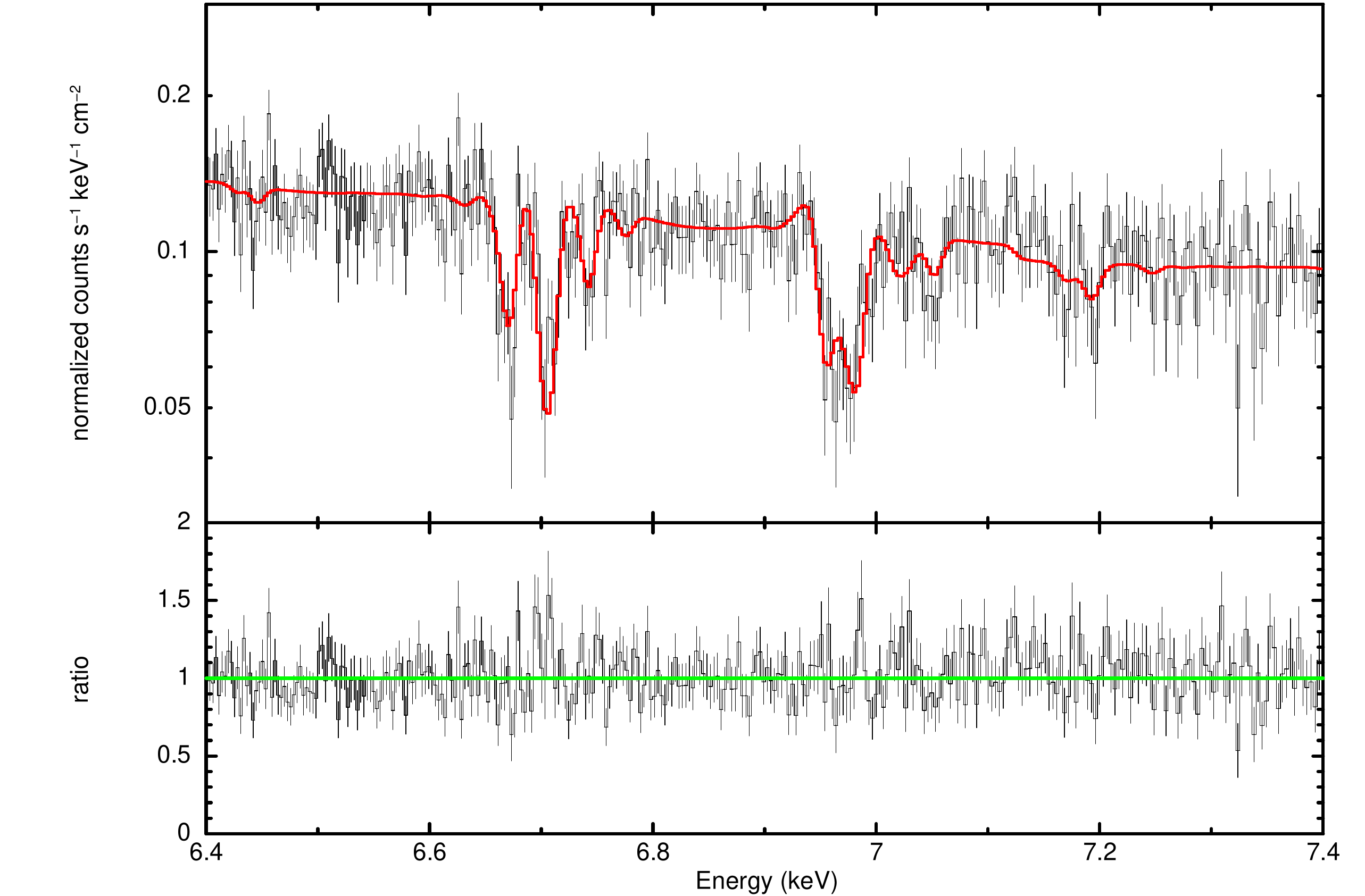}}
\caption{\chandra HETGS 3rd-order spectrum of GRS 1915+105 from \cite{Miller2016a}. The data show strong absorption lines with complex energy structure; the model includes four distinct ionization/velocity zones with associated re-emission components. See text for details.
\label{fig:3rdorder}}    
\end{figure}

One of the first additional candidates proposed for an MHD wind was IGR J17091--3624 \cite{King2012}, who found evidence of absorption lines with blueshifts near $10^4$ km s$^{-1}$. Though thermal driving was possible in this source, the wind speed was better matched to escape conditions at smaller radii in the disk. But advances in modeling have also led to more nuanced views of previously-observed winds. \cite{Miller2015} and \cite{Miller2016a} perfomed detailed photoionization modeling of accretion disk winds in GRO J1655--40, H1743--322, GRS 1915+105, and 4U 1630--47, extending their analysis to the higher-resolution 3rd order \chandra HETGS spectra (see Figure \ref{fig:3rdorder}) and including not only multiple ionization/velocity zones but also photoionized emission (which is expected without fine-tuning of the wind geometry). If the wind structure is rotating, the absorption lines may be narrow but the emission lines should be broadened; \cite{Miller2015} included this effect by blurring the re-emission component. These works found that the required emission line blurring was consistent with Keplerian velocity at the location of the wind inferred via photoionization analysis. Again, these results indicate that the hot winds are not only rotating but also found at small radii (suggesting MHD driving). \cite{Trueba2019} reanalyzed 6 \chandra HETGS observations of 4U 1630--47 and found a similar result, though they also found---like \cite{Neilsen2012b}---that the winds might be best described as hybrid thermal/magnetic outflows. More indirectly, \cite{Hartley2002} used the lack of a correlation between source luminosity and wind activity in two high-state CVs to argue that a non-radiative factor (e.g., MHD processes) might dominate the wind driving.

One point is worth clarifying before going forward: with the exception of observations at high luminosity as described above, thermally-driven winds cannot be launched from small radii. Hot winds found at small radii and luminosity must therefore be driven by MHD processes. But these act throughout the disk, so MHD winds can in principle be launched from or extend to large radii: the detection of wind absorption near the Compton radius is consistent with a thermal wind but cannot in and of itself rule out a magnetic wind.

This distinction is especially relevant for the application of self-similar MHD wind models to X-ray spectra of black holes. \cite{Fukumura2010a} adapted the self-similar wind solution of \cite{Contopoulos1994} to predict the ionization state and spectrum of MHD winds in AGN. A key feature of the model is that rather than being treated as a shell for the purposes of ionization modeling, the wind originates at small radii and extends over many orders of magnitude in radius (with corresponding changes in the density, velocity, and ionization of the gas). These authors applied this model to several supermassive black holes \cite{Fukumura2010b,Fukumura2015} before arguing \cite{Fukumura2017} that it also explained the dense MHD wind in GRO J1655--40 described above and shown in Figure \ref{fig:metastable}. \cite{Fukumura2021} used the same model to fit spectra of GRO J1655--40, H1743--322, and 4U 1630--47, arguing that the state dependence of winds could be explained by a decrease in density and a steepening in the density profile during harder states. Together, these changes effectively serve to render the wind too ionized to observe in absorption.

\cite{Chakravorty2016,Chakravorty2023} have explored another set of self-similar MHD wind models that are based on a solution presented in \cite{Ferreira1997}. The distinction between these winds and the \cite{Fukumura2010a} models discussed above is that in the latter, the initial density of the wind is a free parameter, while in the former it is an explicit function of the disk scale height and the radial profile of the disk accretion rate. \cite{Chakravorty2016} found that the wind launching efficiency was too low for ``cold" solutions that do not include heating of the accretion disk surface, but that the geometry of ``warm" solutions matched the observed inclination dependence of winds \cite{Ponti2012}.

Despite the in-depth observational and theoretical work behind the results above, there is still controversy about evidence for MHD winds in black hole X-ray binaries (especially GRO J1655--40). Based on X-ray spectra and an unusual optical/infrared excess, \cite{Neilsen2016,Shidatsu2016} argued that the dense wind discovered by \cite{Miller2006a} was actually a Compton thick super-Eddington outflow, and while there are still questions about the geometry and radiative transfer of the wind and the OIR excess, this would imply that the intrinsic luminosity of the source must have been much higher. This distinction has important consequences because the metastable $2s^22p_{3/2}$ state can be populated via UV photoexcitation\footnote{Prior studies \cite{Miller2008,Fukumura2017} expressly neglected this effect because the \textit{apparent} Eddington ratio of GRO J1655--40 is low.} as well as collisions \cite[][see also \citenum{Mauche2003,Mauche2004,Mao2017}]{Tomaru2022}. Self-consistently accounting for the higher intrinsic luminosity and SED and allowing the metastable level to be populated by cascades from radiative excitation lowers the required density in the wind by more than an order of magnitude \cite{Tomaru2022} (pushing the inferred launch radius to $\sim10^{11}$ cm). In addition, \cite{Tomaru2022} further argued that the absorption line spectrum in GRO J1655--40 was consistent with an optically thick thermal wind model based on radiation hydrodynamic simulations (this model is shown in Figure \ref{fig:metastable}, including a close-up of the relevant Fe\,{\sc xxii} lines). \vspace{3mm}

Thus it appears that wind absorption line spectra in black hole X-ray binaries can be fit with models implying both small \cite{Miller2015,Miller2016b} and large \cite{Tomaru2022} launch radii, and it seems we are left in a bit of a pickle. But to the extent that our dilemma is the culmination of decades of advances in X-ray data, spectral analysis, and numerical simulations, there is still room for optimism. Will more and more realistic simulations of thermal and magnetic accretion disk winds reveal observable differences between them? The most promising development on the horizon appears to be higher spectral resolution, where the velocity profile of winds may be an important distinguishing factor (see Section \ref{sec:future}).

\subsection{Neutron Stars}\label{subsec:windsNS}

Having reviewed the basics of disk atmospheres and disk winds in the context of black hole XRBs we now turn to neutron star systems. Their accretion properties are overall very similar, which is not surprising because their gravitational potentials are not too different. Nevertheless, neutron stars are fundamentally different objects than black holes, given that they have a solid surface (versus an event horizon) and a magnetic field anchored to it. Both these properties can impact the surrounding accretion flow, and outflows, in various ways. 

Observationally, thermal radiation from the hot neutron star surface, or a boundary/spreading layer where matter from the disk splashes into the star, provides an additional emission component in X-ray spectra. Physically, this additional reservoir of soft X-ray photons can Compton-cool the corona \cite{burke2017}, illuminate the accretion disk \cite{ballantyne2004} and perhaps also affect the thermal stability criteria of the disk (Section~\ref{sec:mdot}). 

The presence of a surface magnetic field can also have a significant impact. The inner disk may be truncated and plasma channeled onto the magnetic poles of the neutron star so that it manifests itself as an X-ray pulsar \cite{Bildsten1997,Wijnands1998}. The magnitude of this effect depends both on the pressure exerted by the accretion flow (which scales with the mass-accretion rate), and the strength of the surface magnetic field. Neutron stars in LMXBs typically have weak magnetic fields ($\lesssim 10^9$~G), apart from a few exceptions where the magnetic field is stronger \cite[$\simeq 10^{10-12}~G$;][]{Bahramian2022}. X-ray pulsars in HMXBs, on the other hand, have strong magnetic fields ($\simeq 10^{12-13}$~G). The stellar magnetic field may also interact with that of the accretion disk and thereby impact the production of both jets and disk winds, where again the strength of the surface magnetic field determines the magnitude of the effect \cite{Romanova2009,Parfrey2017,Vandeneijnden2021}. 

Other differences between neutron stars and black holes that can affect their accretion and ejection properties are that neutron stars, by definition, are less massive than black holes \cite{ozel2016}. In the context of disk winds, for instance, mass is one of the parameters that determine where in the disk a thermal wind can be launched (cf. Eq.~\ref{eq:Rc}). Moreover, neutron stars appear to spin (much) slower than black holes in XRBs\footnote{The fastest rotating neutron star known in a XRB has a spin frequency of 599~Hz \cite{disalvo2022}, which translates into a dimensionless spin parameter of $a\simeq0.3$ \cite{degenaar2015}.}. Finally, likely due to observational biases, neutron star LMXBs are clustered towards smaller orbital periods than their black hole analogues \cite{Bahramian2022}. 
In reviewing our current observational knowledge of disks and winds in neutron star XRBs, we will draw parallels with black hole systems to gauge where their different properties come into play. Table~\ref{tab:winds} gives an overview of the properties of the disk atmospheres and winds observed in neutron star (and black hole) XRBs.

\subsubsection{Observations of disk atmospheres and winds in neutron star systems}
For nearly two dozen neutron star XRBs ($\simeq10\%$ of the currently known population), ionized absorbers have been detected in X-ray spectra (Table~\ref{tab:winds}). Many of these systems are known to be viewed at high inclination, as evidenced by the detection of eclipses and/or dips (resulting from the central X-ray source being obscured by either the companion star or the accretion stream, respectively). For several neutron stars, the X-ray aborption lines are blue-shifted, showing that plasma is outflowing with velocities in the range of $\simeq(1-30) \times 10^{3}$~km~s$^{-1}$ ($\simeq 0.004-0.09$c; Table~\ref{tab:winds}). 
Similar to black hole XRBs, there are some neutron star systems that show wind signatures at other wavelengths but not in X-rays. Examples are the very bright source Sco X-1 \cite[wind feature in the nIR;][]{Bandyopadhyay1999}, and X2127+119 in the globular cluster source M15 \cite[wind feature in the UV;][]{Ioannou2003}).

Although disk winds have so far been detected in only a modest fraction ($\simeq$10\%) of neutron star LMXBs, it is likely that many disk winds go undetected, e.g., due to inclination or ionization effects (Section~\ref{sec:mdot}). Indeed, there is indirect evidence of (highly) non-conservative mass transfer in several neutron star LMXBs, as inferred from the accelerated orbital decay of X-ray pulsars \cite{Marino2019,disalvo2022} and modeling the disk spectral energy distribution in other systems \cite{Ponti2017,HS2019}. In black hole systems there is also indirect evidence for non-conservative mass transfer, as inferred from transient decay light curves \cite{Tetarenko2018}, or modeling the outburst properties \cite{Ziolkowski2018}. Disk winds are an attractive explanation for the severe mass loss that these systems must be experiencing \cite{Dubus2019}. 

\begin{figure}[t]
\centerline{\includegraphics[width=0.85\textwidth,angle=0,clip=true,trim=10 10 10 0]{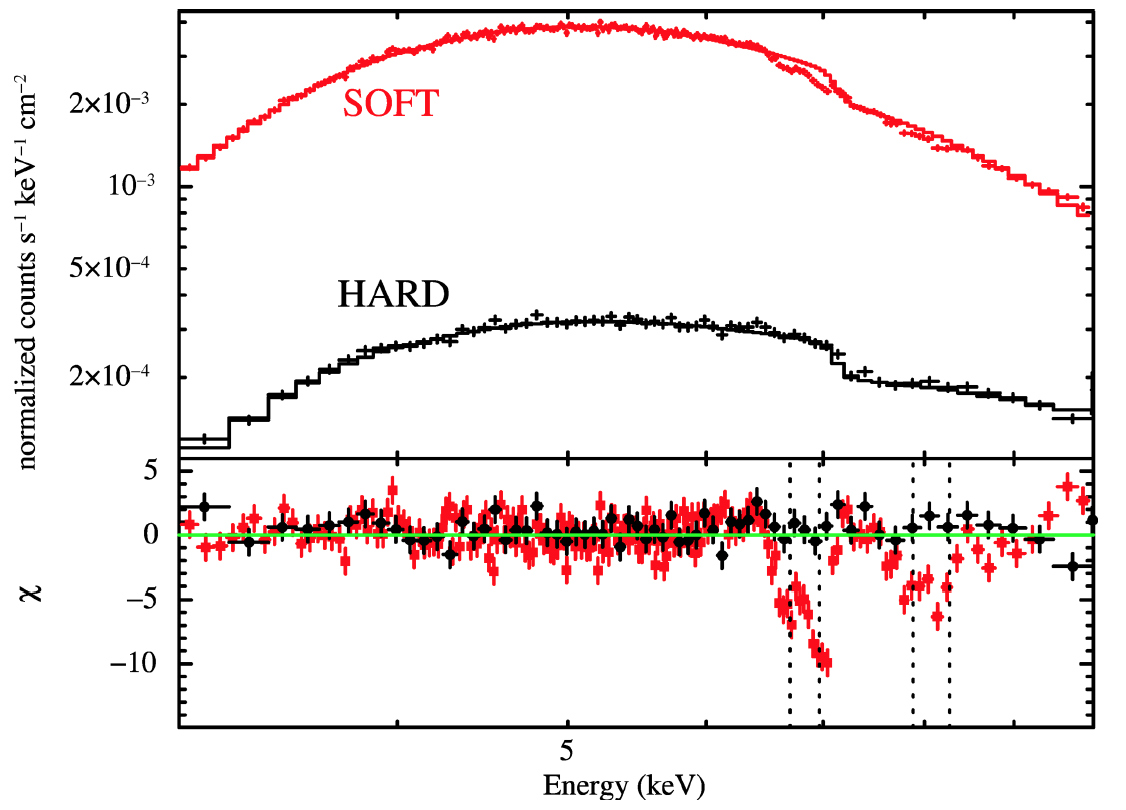}}
\caption{Comparison between the XMM-Newton (CCD) spectra of the neutron star LMXB AX J1745.6--2901 taken in hard (black) and soft (red) states. This clear state-dependence of the observability of the ionized absorber is similar to the behavior seen in black holes. Figure from \cite{Ponti2015}. 
\label{fig:NSdipper}}  
\end{figure}

Similar to the phenomenology seen in black hole systems, the observability of the ionized absorbers in neutron star LMXBs shows a dependence on accretion state. This is, for instance, clearly demonstrated by studies of three high-inclination eclipsing (transient) neutron star low-mass ray binaries across different states: MXB 1659--28 \cite{Ponti2019}, AX J1745.6--2901 \cite[see Figure \ref{fig:NSdipper};][]{Ponti2015}, EXO 0748--676 \cite{Ponti2014}. In all cases, the absorber is clearly seen during soft states, but not during hard states. For the former, the lines are suggested to be at rest (velocities $\lesssim 200$~km~s$^{-1}$), hence indicating the absorber is a static ionised atmosphere rather than an outflowing wind. Indeed, for this system the location of the absorbing plasma was inferred to be (well) within the Compton radius, hence eliminating the possibility of a thermal wind \cite{Ponti2019}. For the other two sources no velocity shifts were detected at CCD energy resolution, which sets an upper limit on the velocity of the absorbing plasma of $\lesssim 10^3$~km~s$^{-1}$. This does not rule out a (slow) wind. 

\begin{table}
\centering
\begin{tabular}{llcccc}
\hline
\hline
Source & $P_\textrm{orb}$ & $N_{\mathrm{Gal}}^{\mathrm{H}}$ & $v_{\mathrm{out}}$ & Class & References\\
&  & ($10^{21}$~cm$^{-2}$)    & ($10^3$~km$^{-2}$) &    &   \\
\hline
\multicolumn{6}{c}{\textbf{Neutron star LMXBs}}\\
**1RXS J180408.9--342058$^a$ & ? & 2 & 26  & SL  & \cite{Degenaar2016} \\
**GX340+0$^b$ & ? & 2 &  12  & SL,HE & \cite{Miller2016a} \\ 
**4U 1820-30 & 0.19h & 1.3 &  1.2 &  & \cite{Costantini2012}\\
**IGR J17062--6143 & 0.63h & 1 &  14  & MP  & \cite{Degenaar2017,vandenEijnden2018} \\ 
XB1916--053 & 0.83h & 2 &  atm & D & \cite{Boirin2004,Juett2006,DiazTrigo2006}\\
 & & & & & \cite{Iaria2006,zhang2014,Trueba2020}\\
1A 1744–-361 & 1.62h & 3 &  atm & D & \cite{Gavriil2012} \\
4U 1323--62 & 2.93h & 10 &  ? & D & \cite{Boirin2005,church2005,Balucinska-Church2009} \\
**XTE J1710--281 & 3.28h  &  2 &  0.8 & D,E & \cite{Raman2018} \\ 
EXO~0748$-$676 & 3.82h  & 0.9 &  $\lesssim 1$  & D,E & \cite{DiazTrigo2006,vanPeet2009,Ponti2014}  \\ 
XB 1254--690 & 3.93h & 2 &  atm & D & \cite{Boirin2003,DiazTrigo2006,DiazTrigo2009,iaria2007} \\
MXB 1659--298 & 7.12h & 2 &  $\lesssim 0.1$  & D,E & \cite{Sidoli2001,DiazTrigo2006,Ponti2018} \\ 
AX~J1745.6$-$2901 & 8.35h & 0.9 &  $\lesssim 1$  & D,E & \cite{Hyodo2009,Ponti2015} \\ 
**IGR J17591--2342 & 8.80h & 10 & 2.8  & MP & \cite{Nowak2019b}\\ 
X1624--490 & 20.9h & 20  &  atm & D & \cite{Parmar2002,DiazTrigo2006,Iaria2007b,Xiang2009} \\
IGR J17480--2446 & 21.3h & 5.5 &  1--3 & MP & \cite{Miller2011} \\
**Swift J1858.6--0814 & 21.3h & 2 &  2 & HE & \cite{Buisson2020}  \\ 
GRO J1744--28 & 11.8d & 10 &  8  & SL,SP & \citep{Degenaar2014a} \\ 
Cir X--1 & 16.6d & 20  &  2 & HE & \cite{Brandt2000,Schulz2002a,Schulz2008}\\
& & & & & \cite{Dai2007,Iaria2007b}\\
GX13+1 & 24.1d & 10 &  out & HE & \cite{Ueda2001,Ueda2004,Sidoli2002}\\
& & & & & \cite{DiazTrigo2012,Madej2014,Dai2014}\\
\hline
\multicolumn{6}{c}{\textbf{Neutron star IMXBs and HMXBs}}\\
Swift J0243.6+6124 & 1.2d & 7&  $66$  & B,SP,HE & \cite{vandenEijnden2019} \\ 
Her~X-1 & 1.7d & 0.15 &  $0.2-1$  & IMXB,SP & \citep{Kosec2020} \\ 
1A 0535+262 & 111d & 4.5 &  $1.5-3.0$  & B & \citep{Reynolds2010} \\ 
\hline
\multicolumn{6}{c}{\textbf{Black hole LMXBs}}\\
4U1630--47 & ? & 17 &  0.3 &  & \cite{kubota2007,DiazTrigo2013,DiazTrigo2014}\\
& & & & & \cite{King2013,King2014,Neilsen2014} \\
H1743--322 & ? & 6.9 &  0.3--0.7 &  & \cite{Miller2006a,King2012} \\
**EXO 1846-031 & ?  &  1.4 &  0.3-18 &  & \cite{Wang2021} \\
**MAXI J1803-298$^c$   &  ? & 2.5  &  ? & D & \cite{Miller2021} \\
**MAXI J1348-630  & ?  & 15  &  10 & SL & \cite{Wu2023}\\  
**MAXI J1631-479   &  ? &  17 &  21 & SL & \cite{Xu2020} \\
XTE J1650--500 & 7.63h  & 0.8 &  0.5 &  & \cite{Miller2002b,Miller2004} \\
MAXI J1305-–704  & 9.74h  & 0.2 &  -9$^d$ &  & \cite{Shidatsu2019,Miller2014b} \\
**4U 1543-47  & 1.1d  & 3.4  &  ? & SL & \cite{Prabhakar2023} \\
GX339--4  & 1.8d & 3.9 &  0.05--0.5 &  & \cite{Miller2004,Juett2006} \\
GROJ1655--40 & 2.6d & 5.1 &  0.5 &  & \cite{Ueda1998,Yamaoka2001,Miller2006b,Miller2008}\\
& & & & & \cite{Netzer2006,DiazTrigo2007,Sala2007b}\\
& & & & &\cite{Kallman2009,Luketic2010,Neilsen2012a}  \\
IGR J17091--3624 & $>$4 d & 5.8 &  15 & SL &  \cite{King2012} \\
**V404 Cyg  &  6.5d & 6.4  & 1.5--3 &  & \cite{Munoz-Darias2022} \\
GRS 1915+105  & 33.5d & 14 &  0.3--1.4 &  & \cite{Kotani2000a,Lee2002,Martocchia2006,Ueda2009}\\
& & & & & \cite{Ueda2010,Neilsen2009b,Neilsen2011,Neilsen2012b,Neilsen2018}\\
\hline
\end{tabular}
\caption{List of disk atmosphere and wind detections in neutron star and black hole X-ray binaries. This is an update of \cite{DiazTrigo2016}; winds reported since then are indicated by **. The quoted absorption column densities ($N_{\mathrm{Gal}}^{\mathrm{H}}$) are solely meant to be indicative and taken from \cite{HI4PI2016}. Upper limits for outflow velocities typically mean that no gratings data were available. Source classification: B=BeXRB, D=dipper, E=eclipser, MP/SP=accreting millisecond/slow X-ray pulsar, HE=high-Eddington source, SL=outflow inferred based on a single significant line detection.\\
$^a$ Proposed $P_{\mathrm{orb}} \lesssim$3~hr \cite{Baglio2016,Degenaar2016}.\\
$^b$ GX340+0 is a Z source; in all Z sources where it is known, $P_{\rm orb}\gtrsim21$~hr \cite{Bahramian2022}.\\
$^c$ Based on the recurrence time of the dips, $P_{\rm orb}$ is expected to be $\simeq7$h \cite{Xu2021}.\\
$^d$ The minus sign indicates an inflow rather than an outflow.
} 
\label{tab:winds}
\end{table}

Whereas the above studies point towards a state dependence for detecting the ionized absorber in X-rays, there are a few (tentative) X-ray wind signatures found for neutron stars during hard states. This includes two LMXB pulsars, IGR J17062--6143 \cite{Degenaar2017,vandenEijnden2018} and IGR J17591--2342 \cite{Nowak2019b}. The former accretes at a very low Eddington ratio ($\simeq0.1$\%) and has a very compact orbit (0.6~hr), hence its disk wind may be driven by magnetic processes (see Section~\ref{subsubsec:nsdriving}). The latter has a wider orbit (8.8~hr) and accreted at only $\simeq1$\% of the Eddington rate during its outburst \cite{Kuiper2020}. During this time it was in a hard X-ray spectral state \cite{Kuiper2020,Manca2023} and it displayed (unusually luminous) radio jet emission \cite{Russell2018}.

\subsubsection{Wind driving mechanisms for neutron star systems}\label{subsubsec:nsdriving}
Accreting neutron stars can launch disk winds via the same physical mechanisms as discussed for black holes (i.e., radiative, thermal or magnetic processes). In neutron star systems there are, however, additional means to launch disk winds. Firstly, winds may be driven by the interaction of the disk and stellar magnetic field \cite{Romanova2009}. Secondly, many neutron stars display thermonuclear bursts, short but intense flashes of Eddington-limited radiation \cite{Galloway2008} that interact with the accretion flow and can potentially also launch a temporary disk wind \cite{Degenaar2018}. 

\begin{figure}[t]
\centerline{\includegraphics[width=0.6\textwidth,angle=0,clip=true,trim=0 0 10 0]{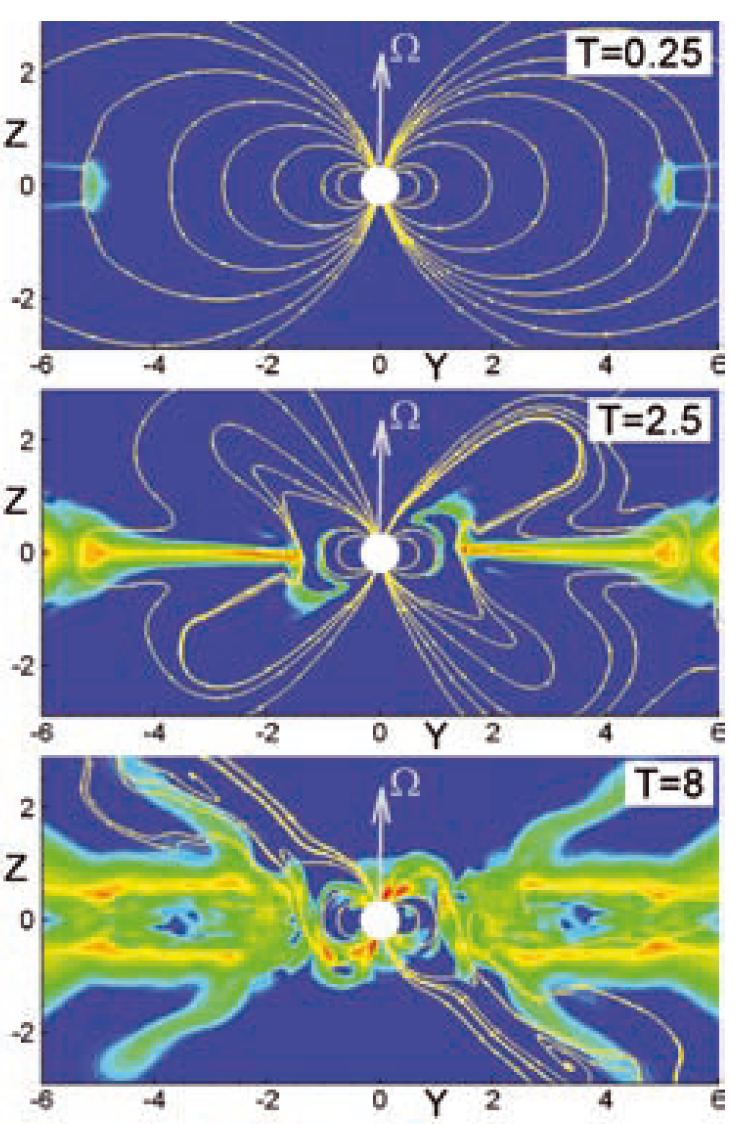}}
\caption{Non-relativistic 3D MHD simulations that show the formation of conical winds from accretion onto a magnetized star \cite{Romanova2009}. Shown are the distribution of matter flux (colors) and sample magnetic field lines at different times (in the YZ plane). The  direction of the stellar magnetic moment and angular velocity of the rotation is indicated by arrows.
\label{fig:romanova}     }  
\end{figure}

\textbf{Disk winds from disk-magnetosphere interaction.} 
Early work on the interaction of a magnetized star with $\alpha$-type disks revealed that outflows may be launched at the disk-magnetosphere boundary \cite{Lovelace1995,Goodson1999,Ustyugova2006,Romanova2009}. In these non-relativistic simulations, conically-shaped winds flow out of the inner disk because in-falling plasma compresses the neutron star magnetosphere, causing the field lines to inflate due to the differential rotation between the disk and the star \cite[][see Figure~\ref{fig:romanova}]{Romanova2009}. Such a conical thin wind with a half-opening angle of $\theta = 30-40 ^{\circ}$ and velocities up to $v = 0.1c$ \cite{Romanova2009} should be detectable with high-resolution spectroscopy in low-inclination systems that harbor a magnetic neutron star. 

More recently, the first suites of GRMHD simulations of accretion onto magnetic neutron stars have been performed. The first calculations for neutron stars with dipolar magnetic fields \cite{Parfrey2017} were further extended to include more complex magnetic fields \cite{Das2022}, motivated by the recent evidence of multi-polar magnetic fields from pulse-profile modeling \cite[e.g.][]{Riley2019}. These GRMHD simulations have not confirmed the presence of the conical wind patterns seen by earlier non-relativistic calculations \cite{Romanova2009,Romanova2012}, but instead find that collimated jet-type outflows are launched. 

Observationally, many of the disk wind detections in neutron star systems are consistent with thermal driving \cite{DiazTrigo2016}. There appear to be, however, interesting exceptions. For instance, there are a few (potential) disk wind detections in neutron star LMXBs that have very short orbital periods that should not allow them to drive thermal winds. Prime examples are 4U1820--30 \cite{Costantini2012} and IGR J17062--6143 \cite{Degenaar2017,vandenEijnden2018}, with orbital periods of 11.4 and 37.4 min, respectively. As their accretion luminosity is well below Eddington, these systems may be, by elimination, good candidates of magnetically-driven winds. In this respect, it is interesting to note that IGR J17062--6143 is an LMXB pulsar \cite{Strohmayer2018} and its accretion disk appears to be truncated by its magnetic field \cite{Degenaar2017}. Several other pulsars with longer orbital periods have recently been reported to show disk winds. This includes the weakly magnetic (LMXB) pulsar IGR J17591--2342 \cite{Nowak2019b} and the strongly magnetic pulsars GRO J1744--28 \cite{Degenaar2014a}, Swift J0243.6--6124 \cite{vandenEijnden2019}, Her X-1 \cite{Kosec2022}. It is interesting to note that the blueshifts of several of these pulsars (IGR J17062--6143, Swift J0243.6--6124, GRO J1744--28) stand out as high compared to those found in other neutron star and black hole XRBs where thermal driving might be the dominant process (see Table~\ref{tab:winds}). This is another reason to suspect that magnetic processes might be involved in driving the disk winds in some neutron star XRBs

\begin{figure}[t]
\centerline{\includegraphics[width=0.85\textwidth,angle=0,clip=true,trim=10 10 10 0]{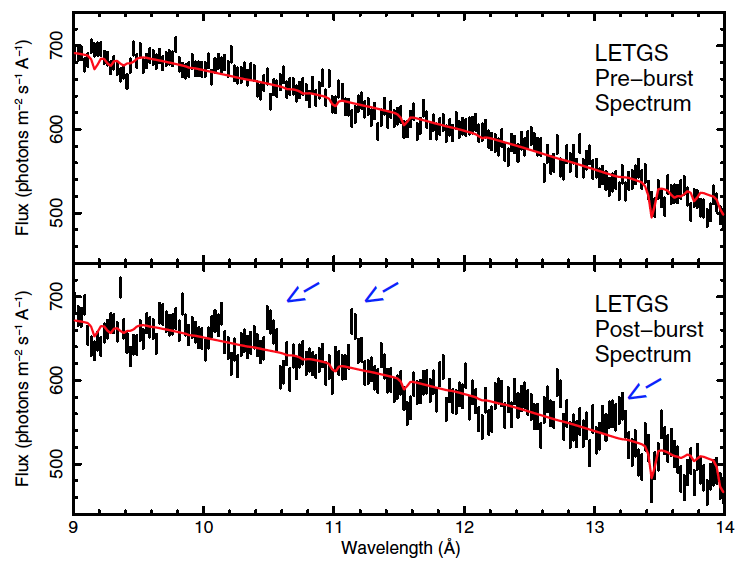}}
\caption{Temporary disk wind detected in the short-period neutron star system SAX J1808.4--3658 \cite{Pinto2014}. The top and bottom panels show the gratings data recorded before and after an energetic thermonuclear burst, respectively. The emergence of narrow, blue-shifted emission lines of Ne\,\textsc{x} and Fe\,\textsc{xxiv} suggest that the intense radiation of the burst may have induced a disk wind \cite{Degenaar2018}.
\label{fig:burstwind}     } 
\end{figure}

\textbf{Burst-driven disk winds.} Another interesting observational development is that there are several neutron star LMXBs for which there is evidence of temporary winds related to bursts \cite[see Figure \ref{fig:burstwind};][]{Degenaar2013,Pinto2014,Strohmayer2019}. 
Their connection to bursts suggests that these disk winds are purely radiatively driven, or perhaps reveal the impact that radiation can have on an existing (magnetic or thermally driven) wind (see Section~\ref{sec:driving}). These examples all concerned particularly energetic bursts, which may have promoted wind detection. Perhaps with future instrumentation (see Section~\ref{sec:future}), the impact of bursts on disk winds can be studied more routinely. It has not been explored in numerical simulations yet what impact bursts may have on driving or enhancing disk winds.

\section{Geometry}\label{sec:geometry}

\begin{svgraybox}
    ...the X-ray luminosity of the source is determined by the wind from the star, i.e., the highest luminosity states coincide with the strongest wind. It also confirms the identification of the disk or accretion structure close to the compact object as the origin for the fluorescent K$\alpha$ line and the blackbody, and it suggests that these features and the associated accretion flow structure (i.e., the accretion disk) are anticorrelated with the wind from the companion. \cite{Kallman2019}
\end{svgraybox}

In addition to outflows, high-resolution spectroscopy of X-ray binaries has proven to be a powerful probe of the geometry of accreting systems, from the innermost stable circular orbits of accretion disks to their ionized outer atmospheres, from the hot accretion columns of white dwarfs to relativistic jets and the clumpy ionized winds from massive stars. In this section, we review a selection of these results.

\subsection{Inner Accretion Flows}

\subsubsection{Inner Disks, Reflection, and Spin}
X-ray spectral diagnostics of accretion disks in strong gravitational fields---whether thermal emission or fluorescent iron lines---rely on the relativistic broadening of emission from the inner flow, a combination of Doppler boosting and gravitational redshifts \cite[e.g.,][]{Reynolds2003,McClintock2006}. This broadening may be used to measure the inner radius of the accretion disk (or the spin of a black hole in cases where this radius corresponds to the innermost stable circular orbit). Because accurate measurements of the underlying continuum are essential, it is tempting to think of reflection and black hole spin measurements as largely the domain of telescopes with low or moderate resolution but significant hard X-ray sensitivity, like \nustar, \rxte, and \integral. But this is inaccurate both historically and in practice: high-resolution spectroscopy can play critical roles in revealing the structure of broad iron lines in X-ray binaries and the spins of black holes. 

First, high-resolution spectra are ideal for disentangling broad and narrow lines, so they can lead to more powerful and robust probes of broad line profiles even if reflection is primarily the domain of larger missions with broader energy ranges \cite{Miller2007}. Indeed, one of the first detections of a relativistically-broadened iron line in an XRB was secured with the help of the \chandra HETGS: \cite{Miller2002a} reported a composite Fe line profile in Cyg X-1, definitively demonstrating the existence of both a narrow and a broadened component. Interestingly, \cite{Torres2005} argue that microcalorimeter observations of such broad lines can constrain the precession of the inner accretion disk. High-resolution spectra have played a similarly important role in uncovering the origin of broad lines in neutron star X-ray binaries. \cite[][for example]{Chiang2016} used grating spectra to show that there was no narrow line emission in Ser X-1, effectively requiring a relativistic origin for the line. Though spin is not a factor because neutron star spacetime is approximately Schwarzschild, these broad lines can therefore probe the physical scale of the inner disk \cite[and provide upper limits on the neutron star radius, e.g.,][and references therein]{Cackett2010,Ludlam2018}.

Second, to the extent that spin and reflection diagnostics are sensitive to the underlying continuum, they are also sensitive to the inferred level of interstellar and ionized absorption (see Section \ref{sec:ism} and discussion in \cite{Nowak2017}). For missions like \nustar and \rxte with little or no soft X-ray sensitivity, indirect constraints on interstellar absorption are degenerate with thermal emission from the accretion disk as well as the low energy tail of any power-law component. High-resolution spectra of absorption edges facilitate direct measurements of the column densities of gas and dust in the ISM \cite[e.g.,][]{Lee2002}. \cite{Nowak2011} performed a detailed analysis of observations of Cyg X-1 with \chandra, \textit{Suzaku}, and \rxte and demonstrated both a composite line at high resolution and the importance of soft X-ray constraints on neutral and ionized absorption for modeling the continuum and broad lines. Again, the presence of coordinated or contemporaneous high-resolution spectra can enhance the robustness of reflection and spin diagnostics.

\subsubsection{Central Engines and Boundary Layers}

Beyond spin and relativistic diagnostics of the inner edge of the accretion disk, high-resolution spectra can provide other insights into the nature of the inner accretion flows around compact objects. 

One powerful probe not available in black hole systems is emission from the boundary layer (or the accretion column in the case of magnetically-driven accretion). \cite{Mukai2017} gives a broad overview of the many subclasses of accreting white dwarf systems, which we will not review in detail here. We do note, however, that  expected line properties generally vary significantly depending on the geometry of the emission region, and therefore we should not be overly quick to generalize between different classes of CVs, e.g., from dwarf novae to polars (where the configuration of the accretion flow is very different).

As in black hole and neutron star X-ray binaries, the spectral behavior of CVs can change with accretion rate. In dwarf novae, for example, high-resolution spectra typically show strong emission lines in quiescence. \xmm RGS spectra of OY Car, for example, indicated a multi-temperature plasma with components at $\sim3$ keV and $\sim7$ keV \cite{Ramsay2001}. \chandra HETGS spectra have revealed similar series of strong lines \cite[e.g.,][]{Szkody2002,Perna2003,Pandel2005}. According to these sources the lines are consistent with an origin in the boundary layer itself, i.e., plasma settling onto the white dwarf from the accretion disk \cite[see][and references therein for a more extensive discussion of this point]{Mukai2017}.

\cite{Hellier2004,Rana2006} discussed the presence of strong fluorescent Fe lines even in quiescent CVs, which likely indicate reflection off the white dwarf surface. \cite{Hayashi2018} performed Monte Carlo simulations of reflection in magnetic CVs to study the dependence of the fluorescence features on parameters like white dwarf mass, spin, viewing angle, abundances, and the specific accretion rate. But there are also notable changes in the emission components in dwarf novae between outburst and quiescence. \cite{Rana2006} found that the Fe\,{\sc xxv} line was significantly broadened during outburst in some systems,  suggesting outflowing material near the white dwarf surface. Indeed, \cite{Mauche2004b} argued that the \chandra LETGS spectrum of SS Cyg in outburst comprised continuum emission from the boundary layer, absorbed and scattered by an outflowing wind.

\begin{figure}[t]
\centerline{\hfill\includegraphics[width=0.475\textwidth]{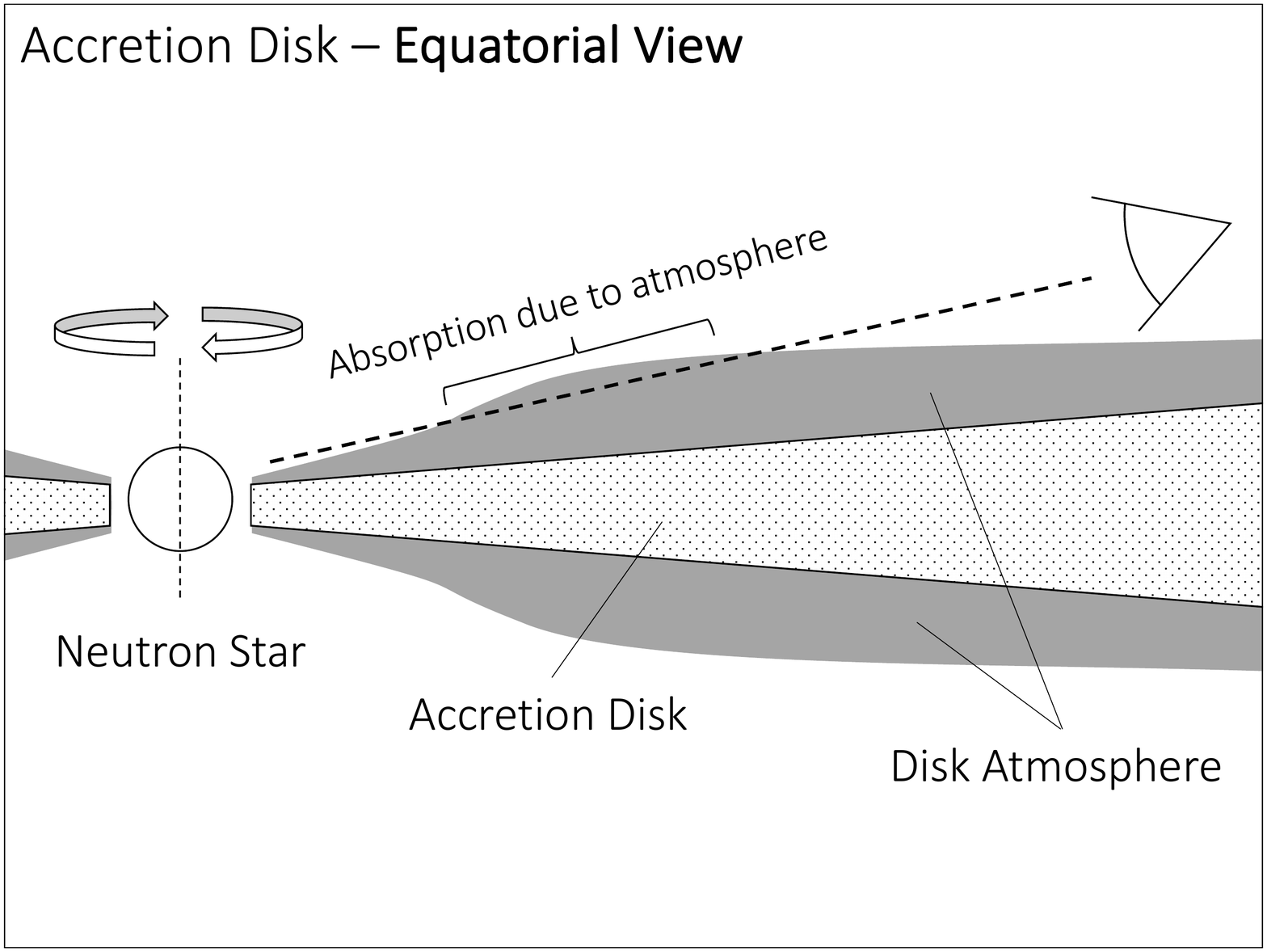}\hfill\includegraphics[width=0.475\textwidth]{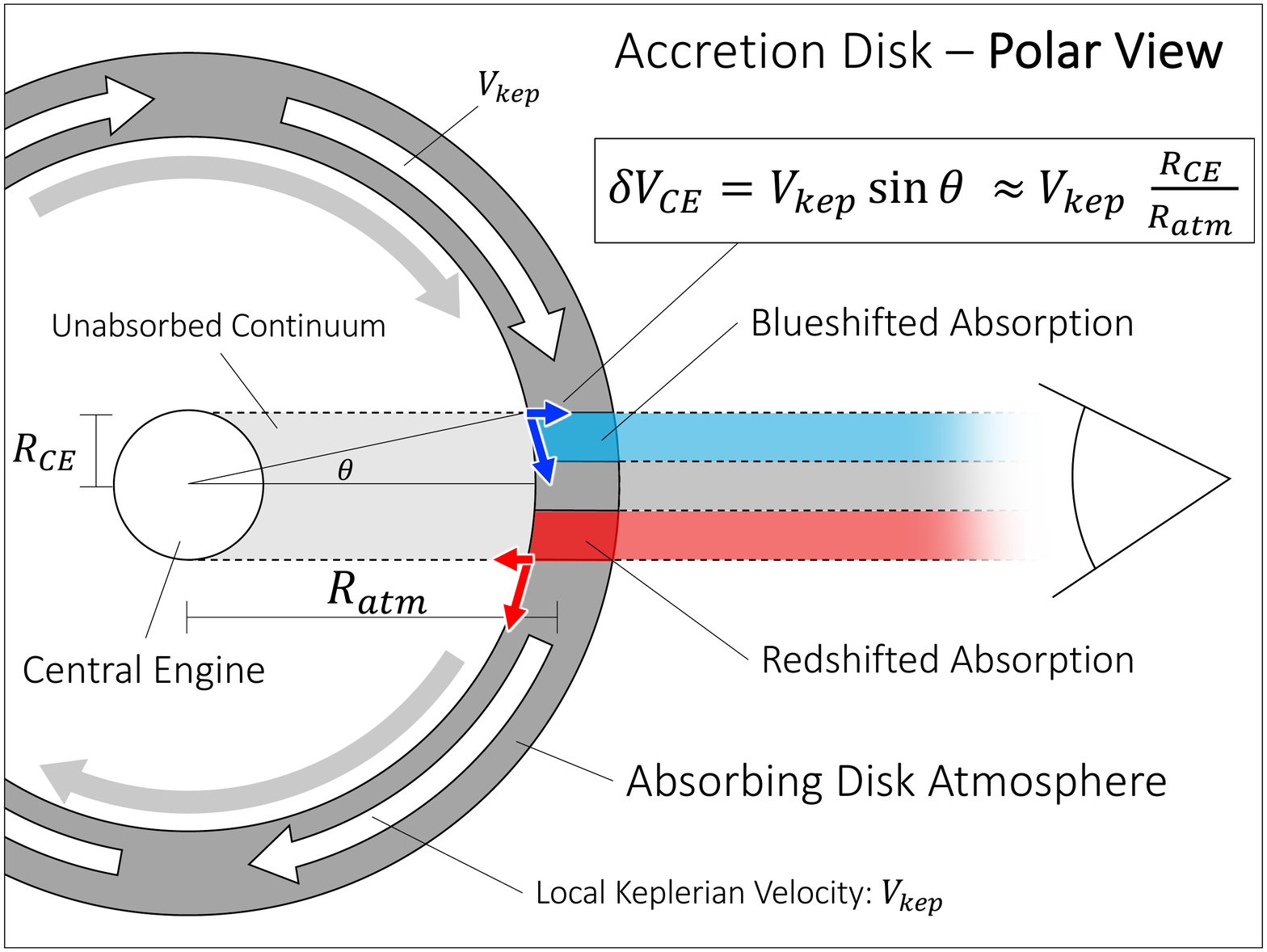}\hfill}
\caption{Cartoon illustrating geometric velocity broadening from \cite{Trueba2022}. A disk atmosphere sufficiently close to the center will exhibit velocity broadening that depends on the physical size of the central engine.\label{fig:cengine}}       % Give a unique label
\end{figure}

From the preceding discussion, it is clear that the physical scale of the inner accretion flow---be it a black hole, neutron star, or white dwarf---is of central importance for accretion diagnostics. While relativistic modeling and ionization analysis are common methods for quantifying the size and location of emitting regions, \cite{Trueba2022} developed a novel high-resolution spectroscopic method for measuring the size scale of the central engine in accreting compact objects using gravitationally-redshifted absorption lines from a rotating disk atmosphere. As shown in Figure \ref{fig:cengine}, the velocity width of these absorption lines will depend in part on the size of the X-ray emitting region. They were able to place $3\sigma$ upper limits of $<90~GM/c^2$ on the size of the central engine in the short-period neutron star LMXB XTE J1710--053. Though it has only been applied in a small number of systems to date, it seems ideal for microcalorimeters on next-generation X-ray missions like \xrism and \athena.

\subsection{Outer Limits}

\subsubsection{Outer Accretion Disks}

Even in cases where the relevant gas is not close enough to the central engine to determine its size scale, emission and absorption lines can provide detailed information about the structure and behavior of accretion disks. One well-established example of outer accretion flow diagnostics comes from dipping LMXBs, where early \xmm and \chandra observations revealed the existence of ionized absorption and emission lines in a number of high-inclination systems \cite[e.g.,][]{Cottam2001a,Cottam2001b,Sidoli2001,Parmar2002,Sidoli2002,Jimenez-Garate2003,Boirin2005}. These observations implied the existence of hot and warm gas extending well above and below the accretion disk midplane at relatively large radii: the extended accretion disk corona (ADC) \cite[e.g.,][]{Church2004,church2005,Iaria2006,Schulz2009}. It is now understood \cite[see][and references therein]{vanPeet2009} that dipping phenomena are caused by transient obscuration of the central X-ray source by structures in the outer accretion disk, such as the disk rim, accretion stream, or the bulge at the disk-stream impact point. \cite{Balucinska-Church2011} used \chandra/\xmm grating spectra of Cyg X-2 to argue that the dips at orbital phase 0.35 could be attributed to structures opposite the accretion stream impact point. More recently, \cite{Psaradaki2018} used X-ray eclipses to map hot gas above the disk in EXO 0748--676. They found two separate ionization components; they interpreted the higher ionization emitter as the extended disk atmosphere, and the cooler gas as clumpy dense gas related to the accretion stream impact itself. Furthermore, to the extent that the ADC or other disk atmosphere are azimuthally asymmetric, any associated emission/absorption features can be expected to vary with orbital phase, a fact used by \cite{Xiang2009} to map the ionization structure and viewing geometry of the ``Big Dipper" 4U 1624--490. 

Similar techniques can be applied to similar benefit to study the structure of disks and their environments directly. For example, high-resolution X-ray spectra have illuminated the behavior of superorbitally-precessing accretion disks in both low-mass and high-mass X-ray binaries. \cite{Jimenez-Garate2002} calculated the expected recombination line emission from an accretion disk irradiated by a central X-ray source. Based on those calculations, \cite{Jimenez-Garate2002} argued that  photoionized emission lines seen in \xmm RGS observations of Her X-1 likely originated in the disk atmosphere or the illuminated face of the secondary star. Drawing on 170 ks of \chandra grating spectra, \cite{Ji2009} inferred cool, dense material distributed throughout the disk in Her X-1. \cite{Kosec2022} divided the emission lines in Her X-1 into three groups, attributed to (1) the outer accretion disk, (2) the disk/magnetosphere boundary, and (3) X-ray reflection from the accretion curtain. This multi-zone structure echoes similar results from \cite{Neilsen2009b} in LMC X-4, where the spectral variability properties of different line groups seen by the \chandra and \xmm gratings could be traced to the inner disk, illuminated outer disk, or the photoionized stellar wind from the massive donor.

There has also been significant interest in modeling and interpreting narrow Fe\,K$\alpha$ lines. Reflection from distant cold gas, for example, is sensitive to the geometry, ionization, dynamics, and abundances in the accretion flow \cite{George1991}\footnote{With upcoming microcalorimeter missions, it may even be possible to use Fe\,K$\alpha$ lines to measure the mass functions of X-ray binaries \cite{DashwoodBrown2022}.}. \cite{Watanabe2003} reported a sensitive measurement of the iron line in GX 301--2, including a detection of the ``Compton shoulder," a faint feature on the red wing of the Fe\,K$\alpha$ line due to backscattering of 6.4 keV photons in a Comptonizing medium. In cases where significant scattering is expected, the Compton shoulder can be used to measure the column density, electron temperature, orientation, abundances, and Compton scattering optical depth \cite[e.g.,][]{Matt2002}. \cite{Torrejon2010} argued that the Fe\,K$\alpha$ line equivalent widths and column densities are consistent with a spherical distribution of reprocessing material around the compact object, though \cite{Tzanavaris2018} concluded the opposite for a smaller sample of X-ray binaries. \cite{Torrejon2010} also reported a decrease of iron line equivalent width with luminosity in X-ray binaries (the so-called ``X-ray Baldwin effect") that they interpreted as the result of photoionization.  Finally, they noted that these lines are ubiquitous in HMXBs but rare in LMXBs, which suggests that illuminated stellar companions and their winds may play a significant role in producing K$\alpha$ complexes in X-ray binaries. Follow-up work on HMXBs by \cite{Gimenez-Garcia2015} used \xmm to confirm both the X-ray Baldwin effect and the contribution of stellar winds.

\subsubsection{Stellar Winds}
It is  clear that stellar winds are important for studies of accretion geometry via fluorescent and recombination lines, but there is also an enormous body of work devoted to the study of stellar winds in their own right. For the purposes of this chapter, we focus our attention on massive stars in XRB systems (particularly Vela X-1 and Cyg X-1, which have received the most attention in recent decades). By way of demonstrating the impact of stellar wind physics for XRB studies, \cite{Gimenez-Garcia2016} noted that the velocity of winds from supergiant donors can have a significant impact on the HMXB class. In other words, stellar winds play an important role in determining the overall behavior of X-ray binaries! Stellar wind physics also has implications for binary evolution scenarios: the difficulty of producing a system like Cyg X-1 (with a rapidly-spinning  21$M_\odot$ black hole in close orbit with a 41$M_\odot$ O star with enhanced He and Fe abundances) requires significant adjustments to prescriptions for wind mass loss rates from the black hole progenitor \cite[][and references therein]{Neijssel2021}.

In Vela X-1, the presence of ionized circumstellar gas was already apparent well before the era of high-resolution X-ray spectroscopy. For example, in 1986, \cite{Nagase1986} used a series of \textit{Tenma} spectra to infer a moderately ionized absorber responsible for an iron edge and emission line. Around the same time, it was becoming clear how complex the structure of such winds would be in a binary system, with accretion wakes and photoionization wakes, as well as a possible accretion stream as sources of large-scale inhomogeneity and clumps on smaller scales \cite{Blondin1990} \cite[see also][]{ElMellah2017,ElMellah2018}. Winds of massive stars were expected to be clumpy due to instabilities associated with line driving \cite{Lucy1970,Lucy1980}, but advances in computation made it possible to simulate the development of shocks in these winds \cite[e.g.,][]{Owocki1988} and to show how they fragment into clumps \cite[e.g.,][and references therein]{Oskinova2012,Sundqvist2013}. \cite{Sako1999} argued that the wind in Vela X-1 had to be inhomogeneous in order to resolve discrepancies between the mass loss rates inferred from optical and X-ray spectra, while \cite{schulz2002c} drew a similar conclusion based on fluorescent line fluxes. Clumps have also been invoked to explain variability on timescales of hours \cite{Furst2010}, absorption dips \cite{Hanke2009}, and associated line variability \cite{Miskovicova2016}. In turn, variability measurements can be used to infer the statistical properties of clumps \cite{ElMellah2020}. 

High-resolution X-ray spectral studies of Cyg X-1 have been similarly informative. A number of early \chandra observations \cite[e.g.,][]{Marshall2001,Schulz2002,Miller2002a} showed strong absorption lines at different orbital phases, while modeling by \cite{Feng2003} indicated that the absorption lines had an extended red wing that was more pronounced in lines from more highly-ionized ions. \cite{Miskovicova2016} found that the line profile actually varies with orbital phase, appearing as symmetric absorption lines at superior conjunction but exhibiting full P-Cygni profiles at inferior conjunction. \cite{Miller2005} used orbital phase comparisons to deduce that the X-ray absorbing portion of the wind must be gravitationally focused: tidal and centrifugal effects produce a distorted, asymmetric wind, as shown by \cite{Friend1982,ElMellah2019} and inferred from optical spectra for Cyg X-1 by \cite{Gies1986}. \cite{Hanke2009} described a \chandra HETGS spectrum with a model including two wind absorption components, which they associated with spherical and focused regions of the wind. \cite{Kallman2019} similarly divided the stellar wind in Cyg X-3 into a ``nebular" component and a ``wind" component \cite[see also][and references therein]{Paerels2000,Suryanarayanan2022}. In Cyg X-1, short-lived dips are commonly observed when the line of sight intersects the densest region of the wind, such that flux-resolved spectra can reveal the ionization structure of clumps in the wind \cite{Hirsch2019}. Careful measurements of line velocities (based on reference energies from an electron beam ion trap)  indicate that the clumps are likely stratified, with denser gas shielding the center and back of the clumps from ionizing radiation. 

In short, as is often the case, the presence of a compact object provides ample opportunities for studying the structure and behavior of stellar winds, while also substantially influencing them. A detailed review of the stellar winds of isolated stars is outside the scope of this chapter, but for a discussion of high-resolution X-ray spectroscopy and the solar wind, we refer the interested reader to Chapter 12 of this volume, by L.\ Gu.

\subsubsection{Jets}
Finally, we turn briefly to spectroscopic constraints on relativistic jets from X-ray binaries. There are several examples of X-ray jets in X-ray binaries in the literature \cite[e.g.,][]{Corbel2002,Steiner2012,Espinasse2020}, but in these cases the X-ray emission arises when jet ejecta interact with the local environment. The best-known example of a spectroscopically-identified jet is SS 433. which exhibits Doppler-shifted lines from its precessing jet. Optical line shifts were attributed to the jet in 1979 \cite{Abell1979,Fabian1979,Milgrom1979}, and Doppler-shifted X-ray emission lines had been associated with the jet within a decade \cite{Watson1986}; \cite[for a review of the X-ray literature, see][and references therein]{Marshall2013}. Currently, the prevailing interpretation is that the X-ray emission lines arise in hot thermal plasma in the jet at distances of $\sim10^{12-13}$ cm from the compact object \cite[e.g.,][]{Brinkmann1991,Kotani1996,Lopez2006,Marshall2013,Medvedev2019}. The advent of high-resolution X-ray spectroscopy has produced a number of insights into the X-ray jet from this important microquasar, including a $\sim5\times$ smaller opening angle in the X-ray than in the optical \cite{Marshall2002} and a large overabundance of Ni \cite[suggesting the donor star may have been enriched by the supernova that created the compact object;][]{Marshall2013}.

It is clear that the advance to microcalorimeters will further advance our understanding of the dynamics and structure of jets. But there are also several results from lower resolution spectra that suggest a similar cause for optimism. For example, \cite{Migliari2002} found hot iron line emission in \chandra CCD spectra of the jet at distances of $>10^{17}$ cm from the compact object \cite[long after adiabatic cooling should have rendered the gas undetectable in X-rays;][]{Marshall2002}. This result implies that there is continued heating of the atomic component of the jet long past the initial acceleration region. More recently, \cite{Middleton2021} identified a time lag in \nustar observations SS 433 that---due to its narrow energy range--could only be attributed to atomic processes. Sensitive observations of this lag with next-generation facilities will reveal its origin in more detail.

More broadly, diagnostics of baryonic jets represent an open question for high-resolution X-ray spectra. In particular, it is notable that there are so \textbf{few} detections of atomic transitions clearly associated with outflowing jets. \cite{DiazTrigo2013b} reported detecting relativistically Doppler-shifted emission lines in 4U 1630--47 using \xmm CCD spectra, though \chandra grating observations of 4U 1630--47 in different accretion states did not show similar emission lines \cite{Neilsen2014}. If confirmed, this would have represented only the second clear case of baryons in a jet from an XRB out of a population of dozens. Next-generation spectra may resolve the question of baryonic jets, but it remains to be seen whether they do so by detecting faint relativistic lines or by revealing baryonic jets to be intrinsically rare.

\section{ISM}\label {sec:ism}

%\begin{svgraybox}
%    One of the most immediate benefits of the high spectral resolution in the soft band is that we can measure the amount of photoelectric absorption almost independent of the intrinsic continuum. \cite{Schulz2002b}
%\end{svgraybox}

In this section, we briefly lay out how high-resolution X-ray spectroscopy can be leveraged to study the ISM. For a more extensive overview, the reader is referred to the recent review of Costantini \& Corrales \cite{Costantini2022}.

Bright X-ray binaries provide an excellent backdrop to study the composition and structure of the ISM at high spectral resolution. Gas along the line of sight to these sources leads to photoelectric edges in X-ray spectra. By comparing the optical depth in these edges to the continuum absorption, it is possible to: (1) measure ISM and stellar abundances \cite[e.g.,][]{Lee2002,Juett2003a}, (2) infer the contributions of interstellar gas vs.\ absorption local to the binary \cite[e.g.,][]{Juett2001,Juett2003a}, (3) quantify the depletion of interstellar gas into dust grains \cite[e.g.,][see below]{Juett2006,Pinto2010}, and (4) map the spatial distribution of cold, warm, and hot interstellar gas \cite[e.g.,][]{Paerels2001,Pinto2010,Gatuzz2018}.

In addition to photoelectric edges, the X-ray spectra of X-ray binaries are imprinted with extinction features from solid dust particles present in the ISM. If X-ray photons interact with electrons inside a dust grain (rather than gas phase atoms), the energy levels can be modified by the effect of neighboring atoms, resulting in an interference pattern referred to as ``X-ray absorption fine structure" (XAFS). The details of such features depend on the energy of electrons and the complexity of the compound, hence allowing us to study the physical properties of interstellar dust such as its chemical composition, the size of the dust grains \cite{Corrales2016} and their crystallinity \cite{Zeegers2017,Zeegers2019}. Bright X-ray binaries provide sufficiently high-quality data to detect these extinction features and can reveal potential differences in the dust properties of the ISM in different Galactic environments \cite{Lee2005,Lee2009}. The chemical composition of interstellar dust appears to be dominated by carbon and silicates, hence the most prominent extinction features are of Si, Mg, O, and Fe (which are the main constituents of silicates). These features have been studied in detail with current X-ray missions \cite{Lee2002,Costantini2012,Pinto2013,Valencic2015,Zeegers2017,Zeegers2019}. Future missions like \athena with higher sensitivity may be able to study similar structures and compositions using edges of less abundant species. 

Another aspect of interstellar dust that is relevant for high-resolution X-ray spectroscopy is dust scattering halos. These halos consist of X-rays from the binary that are scattered into our line of sight by intervening dust screens, and have been observed around a number of XRBs on a variety of angular scales \cite{Audley2001,Ling2009,Tiengo2010,Xiang2011,McCollough2013,Kalemci2018,Jin2019,Sguera2020,Lamer2021}. In addition to practical considerations \cite[e.g., the size of the dust scattering halo compared to the spatial resolution of the instrument][]{Nowak2011,Nowak2017}, through XAFS the structure and distribution of dust grains affect both the intensity profile of dust scattering halos and the shape and depth of photoelectric edges \cite[][and references therein]{Costantini2022}. Future missions that combine sensitivity with spatial and spectral resolution will therefore have the best opportunity to probe the properties of interstellar dust.

\section{Future and Outlook}
\label{sec:future}
Finally, we turn to the future: what can we hope to learn from high-resolution spectroscopy of X-ray binaries with next-generation facilities? As of this writing, there is particular interest in microcalorimeters on \xrism and \athena. The improved sensitivity and spectral resolution of these facilities will be a game changer for all the XRB studies discussed in this chapter. In particular, the ability to compute absorption and emission diagnostics as described in the preceding sections will be greatly enhanced by the reduced time to detect line features, enabling faster constraints on the behavior and geometry of gas in variable systems and more robust detections of weaker lines. Since at the time of writing, the \athena\ mission is in a re-definition stage, the discussion below is focused on \xrism.

The launch of \xrism \cite{Tashiro2020} is anticipated in Spring 2023. It carries a soft X-ray spectrometer, Resolve, which offers a constant $<7$~eV FWHM spectral resolution over the entire 0.3--12 keV bandpass and an effective area of 160 (210) cm$^{-2}$ at 1 (6) keV. 
The spectral resolution of Resolve is a factor 20--40 higher than the CCD instruments on board \chandra and \xmm, while it has substantially increased the collecting area and bandpass over the grating instruments on these missions. Whereas previous studies have shown that hot plasmas in X-ray binaries can have very high velocities, some of the gas may have much smaller characteristic motions of the order of $\simeq$100~km~s$^{-1}$ and hence remain unresolved with current instrumentation.

\begin{figure}[t]
\centerline{\includegraphics[width=0.85\textwidth,angle=0,clip=true,trim=10 10 10 0]{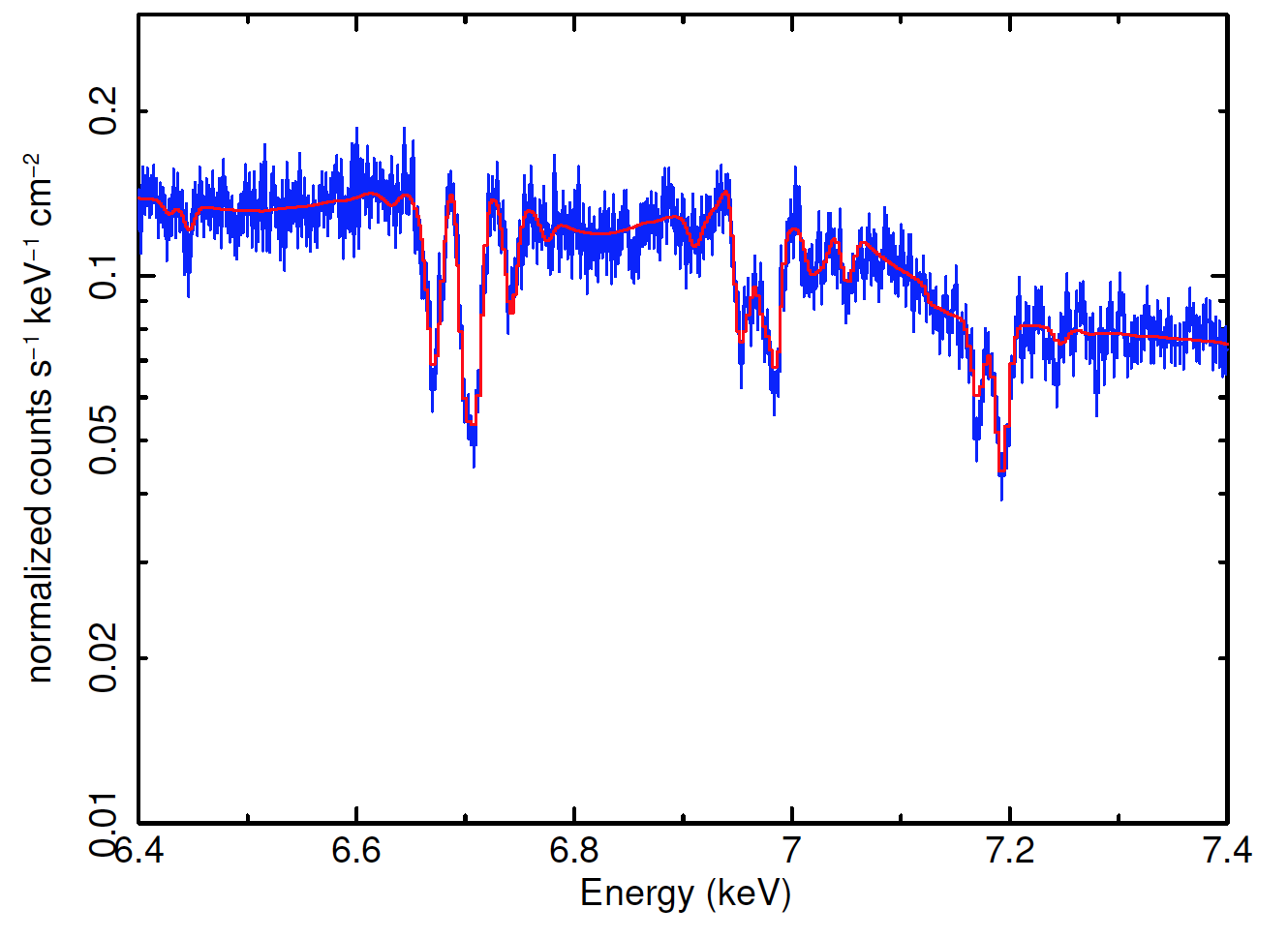}}
\caption{Simulated \xrism/Resolve spectrum of the black hole XRB GRS 1915+105 (50 ks) based on the best \chandra/HETG model \cite{Miller2016b} shown in Figure~\ref{fig:3rdorder}. Reproduced with permission from \cite{XRISM2020}.
\label{fig:xrismsim}     }  
\end{figure}

\textbf{Disk winds.} For X-ray binaries, it remains to be established what the relative importance of thermal and magnetic processes is in driving their disk winds. The increased spectral resolution of new instrumentation will make it possible to separate complexes of lines (Figure~\ref{fig:xrismsim}). This, in turn, will allow us to accurately measure velocities, to separate distinct wind zones, and to measure launching radii. These are all key indicators of the underlying launch mechanism, as discussed in Section \ref{sec:driving}. Also, by virtue of a broad pass band, density-sensitive lines from intermediate charge states become accessible; these enable direct measurements of the density so that the absorption radius can be derived from the photoionization parameter. Wind launch radii can also be measured independently from the velocity broadening of the emission components in lines that have a P-Cygni profile. Furthermore, high-resolution spectrometers may  shed new light on whether or not disk winds are intrinsically transient: if the apparent disappearance in the hard state is due to lower column densities and higher ionization, the increased sensitivity and resolution of \xrism/Resolve compared to current instrumentation may detect hot X-ray winds in hard states.

In this context, it is worth noting that X-ray polarimetry missions will also contribute to advancing studies of XRB disk winds, albeit less directly. For instance, \ixpe has already started to put constraints on the magnetic field topology in disks \cite{Krawczynski2019,Krawczynski2022}, which facilitates further advancement in numerical simulations of disk accretion and any resulting outflows.

\textbf{Geometry}. We may also expect to see large strides in understanding the accretion geometry in X-ray binaries. For instance, the scale height of the outer disk and the extent to which it might be flared, can potentially be revealed by studying narrow emission lines in high-inclination dipping X-ray binaries. 
For stellar winds, the highly increased velocity resolution enabled by \xrism will allow the separation of lines produced close to the
compact object from those excited in the larger stellar wind. Moreover, it will allow a precise detection of velocity shifts as a function of the orbital phase. Improved measures of the ionization, density and temperature of the accreting material will enhance detailed comparisons to hydrodynamic simulations. Furthermore, the increased sensitivity of next-generation missions will facilitate studies of wind structure and clumping in many more systems and on short timescales \cite{Lomaeva2020}. 

\textbf{ISM}. High-resolution X-ray spectroscopy studies of the ISM will also take a leap with new spectrometers. For instance, \xrism provides access to dense environments of the Galaxy ($N_H \simeq 1-10 \times 10^{22}~\mathrm{cm}^{-2}$) near the Galactic Center or molecular clouds, which is currently uncharted territory. \xrism will allow to compare the spectral features of interstellar dust (especially for heavier elements such as Ca, S, and Fe) to material absorption measured in the lab \cite{XRISM2020}.\vspace{5mm}

As detailed in the pages above, the strength of high-resolution spectroscopy lies in the power to illuminate the physical conditions in gas and dust and, accordingly, reveal the physical processes that produce them. In X-ray binaries, these processes---accretion and ejection, ionization, and so on---vary from sub-second timescales to years and decades. Building on existing high-resolution X-ray spectra, next-generation facilities will rewrite our understanding of accretion physics by probing shorter timescales with superior constraints on the ionization and dynamics of gas and dust around compact objects.

\bibliography{book}

\end{document}